\documentclass[%
pra,twocolumn, amsmath,amssymb,superscriptaddress,longbibliography
]{revtex4-2}
\usepackage{xcolor, tensor}
\usepackage{physics}
\usepackage{xparse}
\usepackage{appendix}
\usepackage{graphicx}
\usepackage{dcolumn}
\usepackage{bm}
\usepackage{hyperref}
 \UseRawInputEncoding

\usepackage{bm}
\newcommand{\be}{\begin{equation}}
\newcommand{\ee}{\end{equation}}

\usepackage{bbold}

\begin{document}

\title{Solution of the BEC to BCS Quench in One Dimension}
\date{\today}

\author{Colin Rylands}
\affiliation{SISSA and INFN, via Bonomea 265,  34136 Trieste, Italy}

\author{Pasquale Calabrese}
\affiliation{SISSA and INFN, via Bonomea 265,  34136 Trieste, Italy}
\affiliation{International Centre for Theoretical Physics (ICTP), Strada Costiera 11, 34151 Trieste, Italy}

\author{Bruno Bertini}
\affiliation{School of Physics and Astronomy, University of Nottingham, Nottingham, NG7 2RD, UK}
\affiliation{Centre for the Mathematics and Theoretical Physics of Quantum Non-Equilibrium Systems,
University of Nottingham, Nottingham, NG7 2RD, UK}

\begin{abstract}

A gas of interacting fermions confined in a quasi one-dimensional geometry shows a BEC to BCS crossover upon slowly driving its coupling constant through a confinement-induced resonance. On one side of the crossover the fermions form tightly-bound bosonic molecules behaving as a repulsive Bose gas, while on the other they form Cooper pairs, whose size is much larger than the average inter-particle distance. Here we consider the situation arising when the coupling constant is varied suddenly from the BEC to the BCS value. Namely, we study a BEC-to-BCS {quench}. By exploiting a suitable continuum limit of recently discovered solvable quenches in the Hubbard model, we show that the local stationary state reached at large times after the quench can be determined \emph{exactly} by means of the Quench Action approach. We provide an experimentally-accessible characterisation of the stationary state by computing local pair correlation function as well as the quasi-particle distribution functions. We find that the steady state is increasingly dominated by two particle spin singlet bound states for stronger interaction strength but that bound state formation is inhibited at larger BEC density. The bound state rapidity distribution displays quartic power law decay suggesting a violation of Tan's contact relations.  
\end{abstract}
\date{\today}

\maketitle


Upon changing the strength of its coupling, the same quantum many-body system can pass from a Bose-Einstein condensate (BEC) to a Bardeen-Schrieffer-Cooper (BCS) superconducting state. This astonishing physical phenomenon is known as BEC-to-BCS crossover~\cite{strinati2018the} and the possibility of its occurrence has been debated by theoreticians since the advent of the BCS theory~\cite{bardeen1957theory}. The issue has been finally settled during the first years of the millennium, when the BEC-to-BCS crossover has been realised experimentally in the context of ultracold fermionic gases with attractive interactions~\cite{bloch2008many, inguscio2008ultra}. In these systems, the BEC regime is reached when the fermions form tightly bound bosonic molecules, which turn into overlapping Cooper pairs when transitioning to the BCS phase. In fact, fermionic gases allow for the observation of the entire crossover, also the intermediate situation where the size of the pairs is comparable to the average inter-particle spacing that is known as the unitary regime~\cite{zwerger2011bcs}.

Even though the basic mechanism for the crossover is understood, at least for the  ``balanced gases" where all fermions can form pairs, a complete analytical description for this phenomenon in three dimensions has not been achieved~\cite{zwerger2011bcs}. On the other hand, the seminal works~\cite{fuchs2004exactly, tokatly2004dilute} showed that such an exact description can be attained considering a quasi one-dimensional setting. First, one notes that the full crossover can be observed in a two-component Fermi gas confined in a quasi one-dimensional geometry by driving the coupling constant through a confinement-induced resonance~\cite{olshanii1998atomic, bergeman2003atom} --- which can be thought of as the one-dimensional analogue of a Feshbach resonance~\cite{bergeman2003atom, bloch2008many}. Second, and this is the main insight of Refs.~\cite{fuchs2004exactly, tokatly2004dilute}, one realises that the entire BEC-to-BCS crossover can be described by integrable quantum gases~\cite{guan2013fermi, guan2022new, minguzzi2022strongly}, allowing for a precise and detailed characterisation of many aspects of the crossover~\cite{iida2005exact, wadati2007bcs, hu2007phase, orso2007attractive, heidrich2010bcs, feiguin2012bcs, pecci2021probing}. More specifically, on the attractive side of the resonance the fermionic gas is accurately described by the Gaudin-Yang model~\cite{gaudin1967systeme, yang1967some} of interacting spinful fermions, while on the repulsive side by the Lieb-Liniger model~\cite{lieb1963exact}, which involves bosonic particles.  This surprising change of statistics happens because the two-particle scattering potential of three dimensional fermions confined to one dimension maintains a bound state also for repulsive interactions, which has been characterised both theoretically~\cite{fuchs2004exactly} and experimentally~\cite{moritz2005confinement}. This quasi one-dimensional system is in contrast to a strictly one-dimensional Fermi  gas wherein there is no bound state on the repulsive side.  Because of the bound state in the two-particle scattering potential, on the repulsive side of the resonance fermions form tightly bound bosonic molecules with an effective short-range repulsive interaction~\cite{mora2005four}. 

Given a system displaying BEC-to-BCS crossover an intriguing question is what happens if the change in the coupling connecting the two regimes is performed suddenly rather than adiabatically. 
Namely, if instead of a BEC-to-BCS crossover one considers a BEC-to-BCS quench. As for the case of adiabatic changes, quantum quenches are experimentally realisable in cold atomic gases~\cite{bloch2008many, kinoshita2006quantum, langen2015experimental,schemmer2019generalized, bouchoule2022generalized,malvania2021generalized,
schneider2012fermionic,hackerman2010anomalous}, however, since they drive the system out of equilibrium, they generate a much richer phenomenology. In fact, the study of quantum quenches has led to numerous conceptual breakthroughs on the theoretical understanding of thermalization~\cite{PolkovnikovReview, calabrese2016introduction, VidmarRigol, essler2016quench, doyon2020lecture, bastianello2022introduction, alba2021generalized} and quantum information spreading~\cite{CalabreseCardy, calabrese2016quantum, AlbaCalabrese1,Calabrese2020, bertini2022growth, nahum2017quantum, zhou2020entanglement}.  In spite of this, up to now a quench from BEC to BCS has been considered only in the three dimensional setting~\cite{yuzbashyan2015quantum,
kelly2022resonant}, where the problem can only be studied in a suitable mean-field approximation. The goal of this letter is to fill this gap: In the spirit of Refs.~\cite{fuchs2004exactly, tokatly2004dilute} we show that an exact description of the BEC-to-BCS quench can be obtained by considering the quasi one-dimensional setting. 

\footnotetext[1]{See supplementary material that contains (i) A brief review of the Bethe Ansatz solution of the Gaudin-Yang model; (ii) An explicit calculation of the overlap between the Bethe states and the BEC state~\eqref{eq:BEC}; (iii) The explicit form of the quench action for quenches from the BEC state~\eqref{eq:BEC}; A derivation of $g_2(\infty)$ using Feynman-Hellmann Theorem.}

More specifically, we consider a system of spinful fermions in one dimension which is prepared in the BEC state
\be
\!\!\!\ket{\Psi_0}\!=\!\!  \int_{\mathcal D_{\!\cal N}}\!\!\!\!\!\!\!{\rm d}{\boldsymbol x}\,\, \Psi^\dag_-(x_1)\Psi^\dag_+(x_1)\cdots\Psi^\dag_-(x_{\cal N})\Psi^\dag_+(x_{\cal N})\!\ket{0}\!,
\label{eq:BEC}
\ee
where $\{\Psi^{\dag}_\sigma(x),\Psi_\sigma(x)\}_{\sigma=\pm}$ denote canonical fermionic creation and annihilation operators and we introduced the region $
\mathcal D_\mathcal{N} = \{ {\boldsymbol x}\in \mathbb R^\mathcal{N},\,\,  0\leq x_1 < \ldots < x_\mathcal{N}\leq L\}$ with $L$ designating the volume of the system. For $t>0$ we let the system to evolve according to the attractive Gaudin-Yang Hamiltonian~\cite{guan2013fermi} 
\begin{align}
H_{\rm GY}=& -\sum_{\sigma=\pm} \int_0^{{L}}\!\!{\rm d}x\,\, \Psi^\dag_\sigma(x)\frac{1}{2m}\partial^2_x \Psi_\sigma(x)\notag\\
&+\frac{c}{m} \int_0^{{L}}\!\!{\rm d}x \,\Psi^\dag_+(x)\Psi_+(x)\Psi^\dag_-(x)\Psi_-(x),
\label{eq:Ham}
\end{align}
where $c$, which we assume to be negative throughout the paper, denotes the strength of the contact interaction and from now on we conveniently set $m=1/2$.  Note that $\ket{\Psi_0}$ is not an eigenstate of $H_{\rm GY}$ for any $c$ but does represent the ground state of the quasi one-dimensional system in the limit of infinite repulsion. 

Our goal is to characterise the infinite-time limit of expectation values of local observables 
\begin{equation}
\mathcal O_\infty \equiv\! \lim_{t\to\infty} \lim_{\rm th}\frac{\mel{\Psi_0}{e^{i t H_{\rm GY}} \hat{\mathcal{O}}(x) e^{ -i t H_{\rm GY}}}{\Psi_0}}{\braket{\Psi_0}{\Psi_0}},
\label{eq:statval}
\end{equation}
where $ \lim_{\rm th}$ denotes the thermodynamic limit of $L\to\infty$ with fixed density of particles $\mathcal N/L$, and $\mathcal O(x)$ is a generic local observable acting non-trivially only around the point $x$. We dropped the $x$ dependence because the problem is translational invariant. 

As shown by Gaudin and Yang~\cite{gaudin1967systeme, yang1967some}, the  Hamiltonian~\eqref{eq:Ham} is integrable via \emph{coordinate} Bethe ansatz~\cite{Takahashi, Note1} and possesses an extensive number of local  conservation quantities.  Therefore in the infinite time limit the system will be described by a generalized Gibbs ensemble constructed from this extensive set of conserved charges.   A description of the expectation values \eqref{eq:statval} can be achieved using the Quench Action method~\cite{CauxEssler, Caux}. The main idea is to postulate the existence of a suitable \emph{representative eigenstate} $|\Phi{{\rangle}}$ of the time-evolving Hamiltonian which fulfils 
\begin{equation}
\mathcal O_\infty  = \lim_{\rm th}\frac{\braket*{\Phi}{\hat{\mathcal{O}}(x)|\Phi}}{\braket{\Phi}{\Phi}}.
\label{eq:statval}
\end{equation}
Namely $|\Phi{{\rangle}}$ gives a microcanonical representation of the local stationary state. Crucially, relying on the integrability of the time-evolving Hamiltonian, Ref.~\cite{CauxEssler} has shown that $|\Phi{{\rangle}}$ can be determined as the saddle point of a certain functional integral. To express it explicitly we need to recall some basic facts about the Bethe-Ansatz solution of the Hamiltonian~\eqref{eq:Ham}. The eigenstates of the Hamiltonian are parameterised by the solutions $\boldsymbol  k =\{k_i\}_{i=1}^N$ and $\boldsymbol  \lambda =\{\lambda_\alpha\}_{\alpha=1}^M$ of the  Bethe equations   
\begin{align}
\label{eq:Bethe1}
\prod_{\alpha=1}^M\frac{\lambda_\alpha-{k_j}-ic/2}{\lambda_\alpha-{k_j}+ic/2}&=e^{i k_j {L} },\\
\prod_{i=1}^N\frac{\lambda_\alpha-{k_i}-ic/2}{\lambda_\alpha-{k_i}+ic/2}&=\prod_{\beta\neq\alpha}^M\frac{\lambda_\alpha-\lambda_\beta-ic}{\lambda_\alpha-\lambda_\beta+ic}.
\label{eq:Bethe2}
\end{align}
The parameters $\boldsymbol  k$ and $\boldsymbol  \lambda$ are known as ``rapidities" and can be interpreted as (complex) momenta that, due to the interactions, fulfill complicated quantisation conditions. Rapidities encode all the relevant information about the thermodynamics of the system and, in particular, they specify the expectation values of all local conserved charges~\cite{Takahashi}. 

For large volumes $L$ and finite numbers of particles the solutions of the Bethe equations have a simple structure: each solution can be constructed combining a number of basic building blocks formed by rapidities arranged along regular patterns in the complex plane~\cite{Takahashi}. These patterns, called ``strings", describe bound states of physical particles and spins and are specified by a single real number, which can be interpreted as their momentum. Strings can be thought of as different species of elementary particles with different momenta composing a given eigenstate, in analogy with what happens in free theories~\cite{Takahashi}. In particular, in the thermodynamic limit eigenstates are specified by the momentum distributions of the strings~\footnote[2]{The fact that the description in terms of strings survives in the thermodynamic limit is an \emph{assumption}, called string hypothesis, see, e.g,~\cite{Takahashi}.}. 

For the Hamiltonian~\eqref{eq:Ham} we have three kinds of strings: real momenta $k_j$ (corresponding to unbounded fermions); sets of $n$ complex rapidities $\lambda_j$ distributed symmetrically around the real axis (describing a bound state of $n$ spins); and a triple of two complex $k_i$ and a real $\lambda_j$ (describing a bound state of fermions with opposite spin with localization length $\sim 1/|c|$.). Therefore, we introduce the momentum distributions $\rho, \sigma_n$, and $\tilde\sigma$ corresponding to these string types. 

We are now in a position to state the main result of Ref.~\cite{CauxEssler}: the momentum distributions of the representative state $|\Phi{{\rangle}}$ are the saddle point of the following action
\be
\mathcal A[\rho,\sigma_n,\tilde\sigma]=\mathcal E[\rho,\sigma_n,\tilde\sigma]-\mathcal S[\rho,\sigma_n,\tilde\sigma],
\label{eq:QA}
\ee
where $\mathcal S[\cdot]$ counts how many eigenstates with non-zero overlap with the initial state are specified by the same distributions $\rho, \sigma_n, \tilde\sigma$~\cite{Note1}, while 
\be
\mathcal E[\rho,\sigma_n,\tilde\sigma] = - \lim_{\rm th}\frac{2}{L} \log\! \abs{\!\braket*{\Psi_0}{\boldsymbol k, \boldsymbol \mu}},
\label{eq:E}
\ee
is the extensive part of the overlap between the eigenstate specified by $\rho, \sigma_n, \tilde\sigma$, and the initial state. Whilst $\mathcal S[\cdot]$ takes a general form, which has the same structure in all Bethe-Ansatz integrable models, determining $\mathcal E[\cdot]$ is an extremely difficult task because it requires the explicit overlaps between the initial state and the eigenstates of the Hamiltonian. For this reason, a full Quench Action solution has so far been achieved only for special, ``integrable", system-initial state combinations~\cite{PiroliPozsgayVernier, DeNardis,Brockmann, wouters2014quenching, pozsgay2014correlations, MestyanPozsgayTakacsWerner, bertini2016quantum, MestyanBertiniPiroliCalabrese,BertiniSchurichtEssler,BertiniTartagliaCalabrese,PiroliVernierCalabresePozsgay1,PiroliVernierCalabresePozsgay2, piroli2016multiparticle, piroli2016quantum, alba2016the, piroli2016exact, denardis2015relaxation, mestyan2017exact, rylands2019loschmidt,deleeuw2018scalar,deleeuw2020spin,deleeuw2016one,kristjansen2022integrable}.  

Our main result is to show that for the BEC-to-BCS quench $\mathcal E[\cdot]$ can be found explicitly and, therefore, we can characterise exactly the representative eigenstate $|\Phi{{\rangle}}$. To this end, we recover \eqref{eq:BEC} as an appropriate continuum limit of an integrable state of the one-dimensional Hubbard model~\cite{rylands2022integrable}. 

To proceed, let us consider the Hubbard model on a lattice of $L_{\rm lat}$ sites and spacing $a$. Its Hamiltonian reads as 
\begin{align}
\label{HubbardHamiltonian}
H_{\rm H}=&-\sum_{j=1}^{L_{\rm lat}}\sum_{\sigma=\pm}\left( c^\dag_{j a,\sigma}c^{\phantom{\dag}}_{j a+a,\sigma}
+c^\dag_{j a+a,\sigma}c^{\phantom{\dag}}_{j a}\right)\\
&+2N_{\rm H}+U \sum_{j=1}^{L_{\rm lat}} c^\dag_{ja,+}c^{\phantom{\dag}}_{ja,+} c^\dag_{ja,-}c^{\phantom{\dag}}_{ja,-},\notag
\end{align}
where $\{c^\dag_{ja,\alpha}, c_{ja,\alpha}\}$ are canonical spinful fermions on the lattice, we imposed periodic boundary conditions, and we denoted the number operator by 
$
N_{\rm H}= \sum_{j=1}^{L_{\rm lat}} \sum_{\sigma=\pm} c^\dag_{j a,\sigma}c^{\phantom{\dag}}_{j a,\sigma}
$. 
Let us now consider the continuum limit $L_{\rm lat}\to\infty$, $a \to 0$, $U\to0$, with $a L_{\rm lat}=L$ and $U/a=c$ fixed. In this limit, which we denote by $\lim_{\rm cont}$, $H_{\rm H}/a^2$ reduces to \eqref{eq:Ham} upon setting~\cite{EsslerFrahmGohmannKlumper}
\begin{eqnarray}
\Psi^\dag_\sigma(x)=\lim_{\rm cont}\frac{c^\dag_{a j,\sigma}}{\sqrt{a}}\Big|_{x=ja},
\end{eqnarray}
and replacing sums with integrals with the prescription 
$
a\sum_{j=1}^{L_{\rm lat}} f(ja)\to \int_0^{L}{\rm d}x f(x).
$ 
Let us now consider the state 
\begin{align}
\ket*{\tilde \Psi_{0,\rm lat}}=&\prod_{l=1}^{L_{\rm lat}/2} \!\left(\frac{A_{2l} + B_{2l}}{2} \right)\ket{0}\!,\label{eq:inthub}
\end{align}
with 
\begin{align}
A_l = c^\dag_{la,+}c^\dag_{la,-}-c^\dag_{la-a,+}c^\dag_{la-a,-},\\
B_l = c^\dag_{la-a,-}c^\dag_{la,+}-c^\dag_{la-a,+}c^\dag_{la,-}\,.
\end{align}
The state \eqref{eq:inthub} is a particular example of integrable state in the Hubbard model~\cite{rylands2022integrable}. 

\begin{figure}
 \includegraphics[trim=7 0 0 0 ,clip, width=1.08\columnwidth]{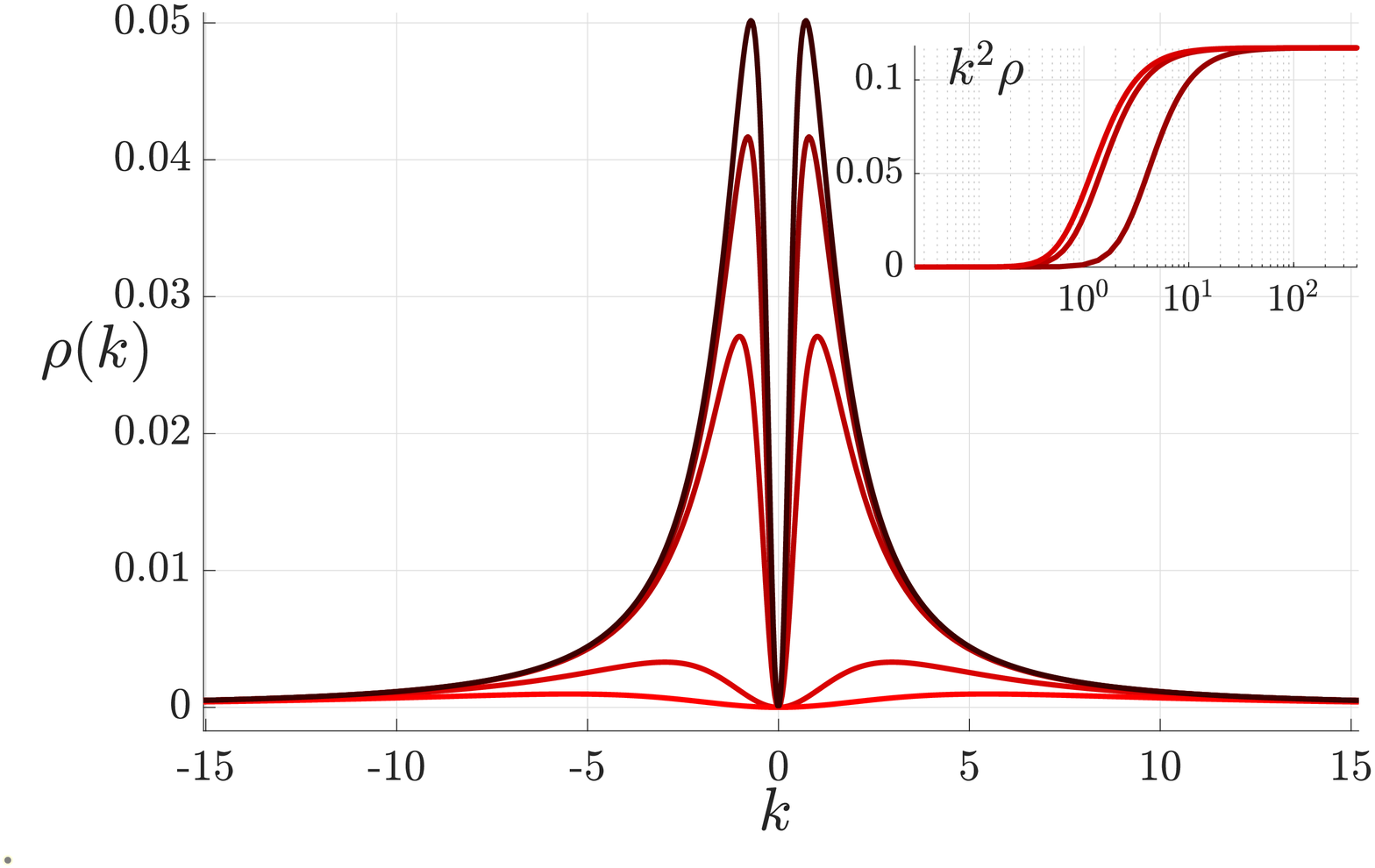} 
 \includegraphics[trim=7 0 0 0 ,clip, width=1.08\columnwidth]{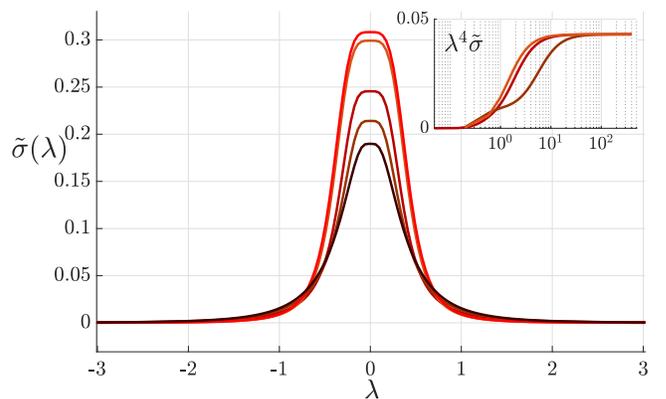}
\caption{\label{fig:RhoDistribution}
Top: The distribution of Bethe rapidities  $\rho(k)$  in the long time steady state for different values of the interaction strength $|c|=.25,.5,1,5,10$ (solid lines, darker to lighter) for fixed density $d=2 \mathcal N/L=0.6$.  Inset: $k^2\rho(k)$ for $|c|=.5,1,5$ indicating a $k^{-2}$ decay of the distribution.  Bottom: $\tilde{\sigma}(\lambda)$ for different values of the interaction strength $|c|=.25,.5,1,5,10$ (solid lines, darker to lighter) at fixed density $d=0.6$.  Inset: $\lambda^4\tilde\sigma(\lambda)$ for $|c|=.5,1,5$ indicating a $\lambda^{-4}$ decay of the distribution. }
\end{figure}
We now relate \eqref{eq:inthub} to \eqref{eq:BEC} reasoning along the lines of Ref.~\cite{brockmann2014overlaps}, where a similar connection has been established between states in the XXZ spin-1/2 chain and the Lieb--Liniger Bose gas. We first adjust the number of particles in the state \eqref{eq:inthub} so that it remains finite and then take the continuum limit. To adjust the particle number we introduce the generators of the so called
``$\eta$ symmetry" of the Hubbard model~\cite{EsslerFrahmGohmannKlumper} 
\be
\eta^z  =\frac{1}{4} L_{\rm lat}- \frac{1}{2} N_{\rm H}, \qquad 
\eta^+ = (\eta^-)^\dag = \sum_{j=1}^{L_{\rm lat}/2} A^\dag_{2j}, 
\ee
which fulfil the $su(2)$ algebra. Using now  
\be
\!\!A^\dag_{2j} \ket{0}=0, \quad A^\dag_{2j} B_{2j}\ket{0}=0,\quad A^\dag_{2j} A_{2j}\ket{0}=2\ket{0},
\ee
it is then simple to prove 
\be
(\eta^+)^{L_{\rm lat}/2-\cal N} \ket*{\tilde \Psi_{0,\rm lat}} = (L_{\rm lat}/2-{\cal N})! \ket{\Psi_{0,\rm lat}},
\ee
where we set 
\be
\!\!\!\!\!\!\ket{\Psi_{0,\rm lat}}\! = \!\!\!\!\sum_{\mathcal D_{{\rm lat}, {\cal N}}}\!\left[\frac{A_{2l_1}\!+\!B_{2l_1}}{2}\right] \ldots \left[\frac{A_{2l_{\cal N}}\!+\!B_{2l_{\cal N}}}{2}\right]\!\! \ket{0}\!,
\ee
and defined $\mathcal D_{{\rm lat}, \mathcal N} = \{ {\boldsymbol i}\in \mathbb N^{\mathcal N},\,\,  0\leq i_1 < \ldots < i_{\mathcal{N}}\leq L_{\rm lat}/2\}$. Finally, noting that in the continuum limit
\be
\frac{A_{2l}+B_{2l}}{2} \mapsto a \Psi^\dag_-(2x)\Psi^\dag_+(2x),
\ee
we see that 
\be
\lim_{\rm cont} \ket{\Psi_{0,\rm lat}}=\frac{1}{2^\mathcal{N}}\ket{\Psi_{0}}\,.
\ee
Using the known formula for the overlaps between \eqref{eq:inthub} and the eigenstates of the Hubbard Hamiltonian (cf. Ref.~\cite{rylands2022integrable}) we can then determine an explicit formula for the overlaps between \eqref{eq:BEC} and the eigenstates of \eqref{eq:Ham}~\cite{Note1}. Plugging into \eqref{eq:E} we finally obtain 
\begin{align}
\mathcal E[\rho,\sigma_n,\tilde\sigma]=& \!\!\!\int \!{\rm d}k\,\rho(k) h(k)+\!\!\int \!\!{\rm d}\lambda\, \tilde\sigma(\lambda)\tilde h(\lambda)\notag\\
&+\!\sum_{n=1}^\infty\! \int \!\!{\rm d}\lambda \sigma_n(\lambda)h_n(\lambda),
\end{align}
where 
\begin{align}\notag
h(x)&=f_1(x)-f_0(x),\quad h_n(x)=\sum_{j=1}^nf_1(x_j)+f_0(x_j),\\
\tilde h(x)&=h(x+ic/2)+h(x-ic/2)+h_1(x),
\end{align}
with $f_n(x)=\log{[(x/c)^2+(n/2)^2]}$ and the sum in $h_n(x)$ is over $x_j=x+i(n+1-2j)c/2$.  Computing then the saddle point of Eq.~\eqref{eq:QA} we obtain a set of integral equations fixing 
the momentum distributions $\rho(k), \tilde\sigma(\lambda)$ and $\sigma_n(\lambda)$ of the steady state \cite{Note1}.

In Fig.~\ref{fig:RhoDistribution} we show some representative examples of $\rho$ and $\tilde\sigma$ for different interactions strengths.  For comparison we recall that for the BEC-BCS crossover only $\tilde{\sigma}$ is non zero~\cite{fuchs2004exactly, tokatly2004dilute} whereas in the quench problem all distributions are non zero for finite $|c|$.  We see, however, that as the interaction strength is increased while particle density, $d=2\mathcal{N}/L$  is held fixed, spectral weight shifts from $\rho$ to $\tilde\sigma$ and therefore the steady state becomes dominated by two particle bound states. Moreover, from the insets we see that the distributions decay in as a power law with a coefficient independent of $|c|$.  Combining this with the $|c|\to0$ limit of the saddle point equations  we find
\be
\lim_{|k|\to\infty}\rho(k)\to\frac{1}{\pi}\Big|\frac{d}{k}\Big |^2,\quad
\lim_{|\lambda|\to\infty}\tilde\sigma(\lambda)\to\frac{1}{\pi}\Big|\frac{d}{\lambda}\Big |^4.
\label{eq:tails}
\ee
  We note that a quartic decay in the rapidity distributions also occurs when quenching the interacting Bose gas, described by the Lieb-Liniger model, from a BEC state~\cite{DeNardis}.
\begin{figure}
 \includegraphics[trim=7 0 0 0 ,clip, width=1.08\columnwidth]{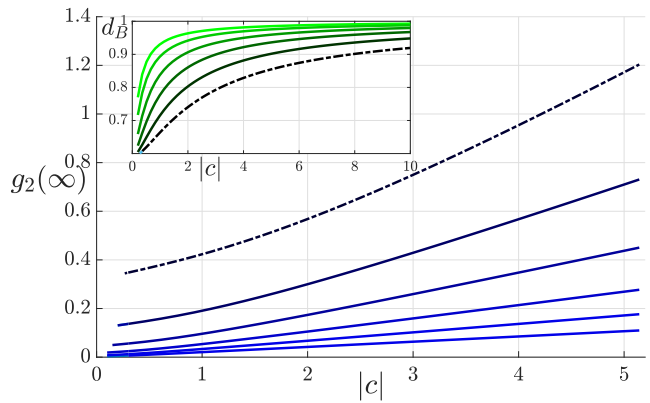} 
\caption{\label{fig:HellFeyn}
Main: The two point correlation function $g_2(\infty)$ in the steady state as a function of $|c|$ for different particle densities $d=0.09,0.14,0.24, 0.37,0.61,1$ (lighter to darker). When we plot $g_2(\infty)/d^2$ against $|c|/d$ all curves collapse onto the $d=1$ line (dot dashed).  Inset: The bound state fraction, $d_B$ as a function of $|c|$ for the same values of the density, darker lines corresponding to higher density.  When plotted against $|c|/d$ all lines collapse onto the $d=1$ curve (dot-dashed). }
\end{figure}

In principle,  the steady-state momentum distributions completely determine all expectation values \eqref{eq:statval}. As for $\mathcal E[\cdot]$, however, finding the explicit form of the functional $\mathcal O_\infty[\cdot]$ requires special operator-system combinations~\cite{mestyan2014short,negro2013on, negro2014on, bertini2016quantum,pozsgay2011local, bastianello2018exact, bastianello2018from}. Here we consider the operator 
\begin{eqnarray}
\label{eq:G2st}
\hat{\mathcal O}(x)=\Psi^\dag_+(x)\Psi_+(x)\Psi^\dag_-(x)\Psi_-(x),
\end{eqnarray}
whose expectation in a stationary state of \eqref{eq:Ham} can be straightforwardly determined via the Feynman-Hellmann theorem~\cite{mestyan2014short, piroli2016quantum, piroli2016multiparticle, bertini2016quantum, rylands2022integrable}.  Note that the expectation value of $\hat{\mathcal{O}}(x)$, typically denoted by $g_2(t)$, is the local pair correlation.  

A direct application of Feynman-Hellmann theorem gives 
\be\label{eq:HellFeyn}
g_{2}(\infty)\!= \!\int \!\!{\rm d}k\, k\, \omega(k)+2 \int \!\!{\rm d}\lambda \,\lambda\,\tilde \mu(\lambda)-\frac{c}{2}\int \!\!{\rm d}\lambda \tilde\sigma(\lambda),
\ee
where $\omega, \tilde \mu$ are determined by solving a set of integral equations~\cite{Note1}.  In Fig.~\ref{fig:HellFeyn} we plot $g_2(\infty)$ as a function of $|c|$ for different values of the density. We see that for small $|c|$ the numerical solutions approach $g_2(\infty)|_{c=0}=d^2/4$ while for large $|c|$, $g_2(\infty)$ displays linear growth and, furthermore, it increases as a function of particle density.  This behaviour can be attributed to the last term of \eqref{eq:HellFeyn} where the integral is merely the number of bound states in the system.  In the inset we plot $d_B=\int \!\!{\rm d}\lambda \tilde\sigma(\lambda)/d$ the fraction of particles forming bound states in the steady state for different densities.  All curves collapse onto the $d=1$  curve (dot-dashed) when plotted against $|c|/d$. In agreement with Fig.~\ref{fig:RhoDistribution} we see that $d_B\to 1$ as $|c|\to\infty$.  Moreover at fixed $|c|$ we note that bound state formation in the steady state is inhibited by  increasing the density. Thus at low densities each initial state molecule is efficiently converted into a two particle bound state.  Upon increasing the density, the presence of nearby molecules causes a competition between the formation of inter-molecular or intra-molecular bound states leading to the creation of unbound particles. 

The pair correlation function can typically be related to the $q^{-4}$ decay of the momentum distribution functions $n_\pm(q)\!=\!\!\int \!\mathrm{d} x\, e^{iqx}\!\expval*{\Psi^\dag_{\pm}(x)\Psi_{\pm}(0)}$ using Tan's universal contact relations~\cite{tan2008energetics, tan2008large, tan2008generalized,barth2011tan, decamp2016high, vignolo2013universal}. In particular, for the Gaudin-Yang gas they read as 
\be
\lim_{|q|\to\infty}q^4n_{\pm}(q)=\mathcal{C}=\frac{2|c|^2}{\pi} g_2.
\label{eq:Tan}
\ee
Importantly, $n_\pm(q)$ are distinct from the rapidity distributions discussed above: the latter describe the real physical excitations of the system while the former is related to the bare fermions $\Psi_\pm(x)$.

The universal relations \eqref{eq:Tan} hold in most stationary states, however, it has been shown that for interaction quenches in the  one-dimensional Bose gas or in the presence of atom loss this breaks down due a $\lambda^{-4}$ tail of the rapidity distribution~\cite{bouchoule2021breakdown}. The asymptotic behaviour~\eqref{eq:tails} then suggests that Tan's relations are violated also for the BEC-to-BCS quench. In fact, a direct application of the arguments of Ref.~\cite{bouchoule2021breakdown} leads us to conjecture that \eqref{eq:Tan} should be modified to 
\be\label{eq:Conjecture}
\mathcal{C}=\frac{2|c|^2}{\pi} g_2(\infty)+\lim_{\lambda\to\infty}\tilde\sigma(\lambda) \lambda^4=\frac{2|c|^2}{\pi} g_2(\infty)+\frac{d^4}{\pi},
\ee
where we have used~\eqref{eq:tails}. Physically, this originates in the truly nonequiibrium nature of the quench, which produces highly excited quasiparticles. This is in stark contrast with the crossover scenario wherein no highly excited quasiparticles are produced and, accordingly, $\tilde{\sigma}$ has support only on a finite interval.

\textit{Discussion.} In this Letter we presented an exact solution of the BEC-to-BCS quench in a quasi one-dimensional Fermi gas. We characterised the stationary state computing exactly its quasiparticle distributions and pair correlation function.  We showed that, differently from the BEC-to-BCS crossover,  the steady state is not comprised solely of two particle bound states but also contains unbound particles. For increasing interaction strength and decreasing density the bound states dominate. Even in this limit, however, the distribution of bound state differs from the crossover displaying power-law tails rather than a sharp cutoff.   Moreover, determining the tails of the rapidity distributions we argued that the stationary state violates Tan's relations and conjectured a modification in Eq.~\eqref{eq:Conjecture}. 

Finding an analytic confirmation of our conjecture is an obvious direction for future research. Crucially, however, our conjecture can also be verified experimentally as both the momentum distributions and the pair correlation distributions are accessible by present-day experiments~\cite{bloch2008many} (as is the  bound state fraction $d_B$~\cite{paintner2019pair, pini2020pair}). Our translational invariant setting can be modelled, for instance, by box-shaped traps like those used in Ref.~\cite{cataldini2021emergent}. The same setup can in principle also allow for a measurement of the quasiparticle distributions using a sequence of transverse and longitudinal expansions followed by time-of-flight measurements. This has been achieved recently for a bosonic gas~\cite{rigol2005fermionization, wilson2020observation} but can be also adapted to the fermionic case~\cite{bolech2012long}.  

{\em Acknowledgments:}
This work has been supported by the Royal Society through the University Research Fellowship No.\ 201101 (BB) and ERC under Consolidator grant number 771536 NEMO (CR and PC).

\bibliographystyle{apsrev4-2}

\bibliography{GYbib}

\begin{thebibliography}{97}%
\makeatletter
\providecommand \@ifxundefined [1]{%
 \@ifx{#1\undefined}
}%
\providecommand \@ifnum [1]{%
 \ifnum #1\expandafter \@firstoftwo
 \else \expandafter \@secondoftwo
 \fi
}%
\providecommand \@ifx [1]{%
 \ifx #1\expandafter \@firstoftwo
 \else \expandafter \@secondoftwo
 \fi
}%
\providecommand \natexlab [1]{#1}%
\providecommand \enquote  [1]{``#1''}%
\providecommand \bibnamefont  [1]{#1}%
\providecommand \bibfnamefont [1]{#1}%
\providecommand \citenamefont [1]{#1}%
\providecommand \href@noop [0]{\@secondoftwo}%
\providecommand \href [0]{\begingroup \@sanitize@url \@href}%
\providecommand \@href[1]{\@@startlink{#1}\@@href}%
\providecommand \@@href[1]{\endgroup#1\@@endlink}%
\providecommand \@sanitize@url [0]{\catcode `\\12\catcode `\$12\catcode
  `\&12\catcode `\#12\catcode `\^12\catcode `\_12\catcode `\%12\relax}%
\providecommand \@@startlink[1]{}%
\providecommand \@@endlink[0]{}%
\providecommand \url  [0]{\begingroup\@sanitize@url \@url }%
\providecommand \@url [1]{\endgroup\@href {#1}{\urlprefix }}%
\providecommand \urlprefix  [0]{URL }%
\providecommand \Eprint [0]{\href }%
\providecommand \doibase [0]{https://doi.org/}%
\providecommand \selectlanguage [0]{\@gobble}%
\providecommand \bibinfo  [0]{\@secondoftwo}%
\providecommand \bibfield  [0]{\@secondoftwo}%
\providecommand \translation [1]{[#1]}%
\providecommand \BibitemOpen [0]{}%
\providecommand \bibitemStop [0]{}%
\providecommand \bibitemNoStop [0]{.\EOS\space}%
\providecommand \EOS [0]{\spacefactor3000\relax}%
\providecommand \BibitemShut  [1]{\csname bibitem#1\endcsname}%
\let\auto@bib@innerbib\@empty
\bibitem [{\citenamefont {Strinati}\ \emph {et~al.}(2018)\citenamefont
  {Strinati}, \citenamefont {Pieri}, \citenamefont {R\"opke}, \citenamefont
  {Schuck},\ and\ \citenamefont {Urban}}]{strinati2018the}%
  \BibitemOpen
  \bibfield  {author} {\bibinfo {author} {\bibfnamefont {G.~C.}\ \bibnamefont
  {Strinati}}, \bibinfo {author} {\bibfnamefont {P.}~\bibnamefont {Pieri}},
  \bibinfo {author} {\bibfnamefont {G.}~\bibnamefont {R\"opke}}, \bibinfo
  {author} {\bibfnamefont {P.}~\bibnamefont {Schuck}},\ and\ \bibinfo {author}
  {\bibfnamefont {M.}~\bibnamefont {Urban}},\ }\href
  {https://doi.org/https://doi.org/10.1016/j.physrep.2018.02.004} {\bibfield
  {journal} {\bibinfo  {journal} {Physics Reports}\ }\textbf {\bibinfo {volume}
  {738}},\ \bibinfo {pages} {1} (\bibinfo {year} {2018})},\ \bibinfo {note}
  {the BCS-BEC crossover: From ultra-cold Fermi gases to nuclear
  systems}\BibitemShut {NoStop}%
\bibitem [{\citenamefont {Bardeen}\ \emph {et~al.}(1957)\citenamefont
  {Bardeen}, \citenamefont {Cooper},\ and\ \citenamefont
  {Schrieffer}}]{bardeen1957theory}%
  \BibitemOpen
  \bibfield  {author} {\bibinfo {author} {\bibfnamefont {J.}~\bibnamefont
  {Bardeen}}, \bibinfo {author} {\bibfnamefont {L.~N.}\ \bibnamefont
  {Cooper}},\ and\ \bibinfo {author} {\bibfnamefont {J.~R.}\ \bibnamefont
  {Schrieffer}},\ }\href {https://doi.org/10.1103/PhysRev.108.1175} {\bibfield
  {journal} {\bibinfo  {journal} {Phys. Rev.}\ }\textbf {\bibinfo {volume}
  {108}},\ \bibinfo {pages} {1175} (\bibinfo {year} {1957})}\BibitemShut
  {NoStop}%
\bibitem [{\citenamefont {Bloch}\ \emph {et~al.}(2008)\citenamefont {Bloch},
  \citenamefont {Dalibard},\ and\ \citenamefont {Zwerger}}]{bloch2008many}%
  \BibitemOpen
  \bibfield  {author} {\bibinfo {author} {\bibfnamefont {I.}~\bibnamefont
  {Bloch}}, \bibinfo {author} {\bibfnamefont {J.}~\bibnamefont {Dalibard}},\
  and\ \bibinfo {author} {\bibfnamefont {W.}~\bibnamefont {Zwerger}},\ }\href
  {https://doi.org/10.1103/RevModPhys.80.885} {\bibfield  {journal} {\bibinfo
  {journal} {Rev. Mod. Phys.}\ }\textbf {\bibinfo {volume} {80}},\ \bibinfo
  {pages} {885} (\bibinfo {year} {2008})}\BibitemShut {NoStop}%
\bibitem [{\citenamefont {Inguscio}\ \emph {et~al.}(2008)\citenamefont
  {Inguscio}, \citenamefont {Ketterle},\ and\ \citenamefont
  {Salomon}}]{inguscio2008ultra}%
  \BibitemOpen
  \bibfield  {author} {\bibinfo {author} {\bibfnamefont {M.}~\bibnamefont
  {Inguscio}}, \bibinfo {author} {\bibfnamefont {W.}~\bibnamefont {Ketterle}},\
  and\ \bibinfo {author} {\bibfnamefont {C.}~\bibnamefont {Salomon}},\ }\href
  {https://books.google.co.uk/books?id=ZNbvAgAAQBAJ} {\emph {\bibinfo {title}
  {Ultra-cold Fermi Gases}}},\ International School of Physics ``Enrico Fermi"\
  (\bibinfo  {publisher} {IOS Press},\ \bibinfo {year} {2008})\BibitemShut
  {NoStop}%
\bibitem [{\citenamefont {Zwerger}(2011)}]{zwerger2011bcs}%
  \BibitemOpen
  \bibfield  {author} {\bibinfo {author} {\bibfnamefont {W.}~\bibnamefont
  {Zwerger}},\ }\href {https://books.google.co.uk/books?id=VjdUd5JbRooC} {\emph
  {\bibinfo {title} {The BCS-BEC Crossover and the Unitary Fermi Gas}}},\
  Lecture Notes in Physics\ (\bibinfo  {publisher} {Springer Berlin
  Heidelberg},\ \bibinfo {year} {2011})\BibitemShut {NoStop}%
\bibitem [{\citenamefont {Fuchs}\ \emph {et~al.}(2004)\citenamefont {Fuchs},
  \citenamefont {Recati},\ and\ \citenamefont {Zwerger}}]{fuchs2004exactly}%
  \BibitemOpen
  \bibfield  {author} {\bibinfo {author} {\bibfnamefont {J.~N.}\ \bibnamefont
  {Fuchs}}, \bibinfo {author} {\bibfnamefont {A.}~\bibnamefont {Recati}},\ and\
  \bibinfo {author} {\bibfnamefont {W.}~\bibnamefont {Zwerger}},\ }\href
  {https://doi.org/10.1103/PhysRevLett.93.090408} {\bibfield  {journal}
  {\bibinfo  {journal} {Phys. Rev. Lett.}\ }\textbf {\bibinfo {volume} {93}},\
  \bibinfo {pages} {090408} (\bibinfo {year} {2004})}\BibitemShut {NoStop}%
\bibitem [{\citenamefont {Tokatly}(2004)}]{tokatly2004dilute}%
  \BibitemOpen
  \bibfield  {author} {\bibinfo {author} {\bibfnamefont {I.~V.}\ \bibnamefont
  {Tokatly}},\ }\href {https://doi.org/10.1103/PhysRevLett.93.090405}
  {\bibfield  {journal} {\bibinfo  {journal} {Phys. Rev. Lett.}\ }\textbf
  {\bibinfo {volume} {93}},\ \bibinfo {pages} {090405} (\bibinfo {year}
  {2004})}\BibitemShut {NoStop}%
\bibitem [{\citenamefont {Olshanii}(1998)}]{olshanii1998atomic}%
  \BibitemOpen
  \bibfield  {author} {\bibinfo {author} {\bibfnamefont {M.}~\bibnamefont
  {Olshanii}},\ }\href {https://doi.org/10.1103/PhysRevLett.81.938} {\bibfield
  {journal} {\bibinfo  {journal} {Phys. Rev. Lett.}\ }\textbf {\bibinfo
  {volume} {81}},\ \bibinfo {pages} {938} (\bibinfo {year} {1998})}\BibitemShut
  {NoStop}%
\bibitem [{\citenamefont {Bergeman}\ \emph {et~al.}(2003)\citenamefont
  {Bergeman}, \citenamefont {Moore},\ and\ \citenamefont
  {Olshanii}}]{bergeman2003atom}%
  \BibitemOpen
  \bibfield  {author} {\bibinfo {author} {\bibfnamefont {T.}~\bibnamefont
  {Bergeman}}, \bibinfo {author} {\bibfnamefont {M.~G.}\ \bibnamefont
  {Moore}},\ and\ \bibinfo {author} {\bibfnamefont {M.}~\bibnamefont
  {Olshanii}},\ }\href {https://doi.org/10.1103/PhysRevLett.91.163201}
  {\bibfield  {journal} {\bibinfo  {journal} {Phys. Rev. Lett.}\ }\textbf
  {\bibinfo {volume} {91}},\ \bibinfo {pages} {163201} (\bibinfo {year}
  {2003})}\BibitemShut {NoStop}%
\bibitem [{\citenamefont {Guan}\ \emph {et~al.}(2013)\citenamefont {Guan},
  \citenamefont {Batchelor},\ and\ \citenamefont {Lee}}]{guan2013fermi}%
  \BibitemOpen
  \bibfield  {author} {\bibinfo {author} {\bibfnamefont {X.-W.}\ \bibnamefont
  {Guan}}, \bibinfo {author} {\bibfnamefont {M.~T.}\ \bibnamefont
  {Batchelor}},\ and\ \bibinfo {author} {\bibfnamefont {C.}~\bibnamefont
  {Lee}},\ }\href {https://doi.org/10.1103/RevModPhys.85.1633} {\bibfield
  {journal} {\bibinfo  {journal} {Rev. Mod. Phys.}\ }\textbf {\bibinfo {volume}
  {85}},\ \bibinfo {pages} {1633} (\bibinfo {year} {2013})}\BibitemShut
  {NoStop}%
\bibitem [{\citenamefont {Guan}\ and\ \citenamefont {He}(2022)}]{guan2022new}%
  \BibitemOpen
  \bibfield  {author} {\bibinfo {author} {\bibfnamefont {X.-W.}\ \bibnamefont
  {Guan}}\ and\ \bibinfo {author} {\bibfnamefont {P.}~\bibnamefont {He}}}
  (\bibinfo {year} {2022}),\ \bibinfo {note} {{arXiv:}2207.01153}\BibitemShut
  {NoStop}%
\bibitem [{\citenamefont {Minguzzi}\ and\ \citenamefont
  {Vignolo}(2022)}]{minguzzi2022strongly}%
  \BibitemOpen
  \bibfield  {author} {\bibinfo {author} {\bibfnamefont {A.}~\bibnamefont
  {Minguzzi}}\ and\ \bibinfo {author} {\bibfnamefont {P.}~\bibnamefont
  {Vignolo}},\ }\Eprint {https://arxiv.org/abs/arXiv:2201.02362}
  {arXiv:2201.02362}  (\bibinfo {year} {2022})\BibitemShut {NoStop}%
\bibitem [{\citenamefont {Iida}\ and\ \citenamefont
  {Wadati}(2005)}]{iida2005exact}%
  \BibitemOpen
  \bibfield  {author} {\bibinfo {author} {\bibfnamefont {T.}~\bibnamefont
  {Iida}}\ and\ \bibinfo {author} {\bibfnamefont {M.}~\bibnamefont {Wadati}},\
  }\href {https://doi.org/10.1143/JPSJ.74.1724} {\bibfield  {journal} {\bibinfo
   {journal} {Journal of the Physical Society of Japan}\ }\textbf {\bibinfo
  {volume} {74}},\ \bibinfo {pages} {1724} (\bibinfo {year}
  {2005})}\BibitemShut {NoStop}%
\bibitem [{\citenamefont {Wadati}\ and\ \citenamefont
  {Iida}(2007)}]{wadati2007bcs}%
  \BibitemOpen
  \bibfield  {author} {\bibinfo {author} {\bibfnamefont {M.}~\bibnamefont
  {Wadati}}\ and\ \bibinfo {author} {\bibfnamefont {T.}~\bibnamefont {Iida}},\
  }\href {https://doi.org/https://doi.org/10.1016/j.physleta.2006.07.068}
  {\bibfield  {journal} {\bibinfo  {journal} {Physics Letters A}\ }\textbf
  {\bibinfo {volume} {360}},\ \bibinfo {pages} {423} (\bibinfo {year}
  {2007})}\BibitemShut {NoStop}%
\bibitem [{\citenamefont {Hu}\ \emph {et~al.}(2007)\citenamefont {Hu},
  \citenamefont {Liu},\ and\ \citenamefont {Drummond}}]{hu2007phase}%
  \BibitemOpen
  \bibfield  {author} {\bibinfo {author} {\bibfnamefont {H.}~\bibnamefont
  {Hu}}, \bibinfo {author} {\bibfnamefont {X.-J.}\ \bibnamefont {Liu}},\ and\
  \bibinfo {author} {\bibfnamefont {P.~D.}\ \bibnamefont {Drummond}},\ }\href
  {https://doi.org/10.1103/PhysRevLett.98.070403} {\bibfield  {journal}
  {\bibinfo  {journal} {Phys. Rev. Lett.}\ }\textbf {\bibinfo {volume} {98}},\
  \bibinfo {pages} {070403} (\bibinfo {year} {2007})}\BibitemShut {NoStop}%
\bibitem [{\citenamefont {Orso}(2007)}]{orso2007attractive}%
  \BibitemOpen
  \bibfield  {author} {\bibinfo {author} {\bibfnamefont {G.}~\bibnamefont
  {Orso}},\ }\href {https://doi.org/10.1103/PhysRevLett.98.070402} {\bibfield
  {journal} {\bibinfo  {journal} {Phys. Rev. Lett.}\ }\textbf {\bibinfo
  {volume} {98}},\ \bibinfo {pages} {070402} (\bibinfo {year}
  {2007})}\BibitemShut {NoStop}%
\bibitem [{\citenamefont {Heidrich-Meisner}\ \emph {et~al.}(2010)\citenamefont
  {Heidrich-Meisner}, \citenamefont {Feiguin}, \citenamefont {Schollw\"ock},\
  and\ \citenamefont {Zwerger}}]{heidrich2010bcs}%
  \BibitemOpen
  \bibfield  {author} {\bibinfo {author} {\bibfnamefont {F.}~\bibnamefont
  {Heidrich-Meisner}}, \bibinfo {author} {\bibfnamefont {A.~E.}\ \bibnamefont
  {Feiguin}}, \bibinfo {author} {\bibfnamefont {U.}~\bibnamefont
  {Schollw\"ock}},\ and\ \bibinfo {author} {\bibfnamefont {W.}~\bibnamefont
  {Zwerger}},\ }\href {https://doi.org/10.1103/PhysRevA.81.023629} {\bibfield
  {journal} {\bibinfo  {journal} {Phys. Rev. A}\ }\textbf {\bibinfo {volume}
  {81}},\ \bibinfo {pages} {023629} (\bibinfo {year} {2010})}\BibitemShut
  {NoStop}%
\bibitem [{\citenamefont {Feiguin}\ \emph {et~al.}(2012)\citenamefont
  {Feiguin}, \citenamefont {Heidrich-Meisner}, \citenamefont {Orso},\ and\
  \citenamefont {Zwerger}}]{feiguin2012bcs}%
  \BibitemOpen
  \bibfield  {author} {\bibinfo {author} {\bibfnamefont {A.}~\bibnamefont
  {Feiguin}}, \bibinfo {author} {\bibfnamefont {F.}~\bibnamefont
  {Heidrich-Meisner}}, \bibinfo {author} {\bibfnamefont {G.}~\bibnamefont
  {Orso}},\ and\ \bibinfo {author} {\bibfnamefont {W.}~\bibnamefont
  {Zwerger}},\ }in\ \href@noop {} {\emph {\bibinfo {booktitle} {The BCS-BEC
  Crossover and the Unitary Fermi Gas}}}\ (\bibinfo  {publisher} {Springer},\
  \bibinfo {year} {2012})\ pp.\ \bibinfo {pages} {503--532}\BibitemShut
  {NoStop}%
\bibitem [{\citenamefont {Pecci}\ \emph {et~al.}(2021)\citenamefont {Pecci},
  \citenamefont {Naldesi}, \citenamefont {Amico},\ and\ \citenamefont
  {Minguzzi}}]{pecci2021probing}%
  \BibitemOpen
  \bibfield  {author} {\bibinfo {author} {\bibfnamefont {G.}~\bibnamefont
  {Pecci}}, \bibinfo {author} {\bibfnamefont {P.}~\bibnamefont {Naldesi}},
  \bibinfo {author} {\bibfnamefont {L.}~\bibnamefont {Amico}},\ and\ \bibinfo
  {author} {\bibfnamefont {A.}~\bibnamefont {Minguzzi}},\ }\href
  {https://doi.org/10.1103/PhysRevResearch.3.L032064} {\bibfield  {journal}
  {\bibinfo  {journal} {Phys. Rev. Research}\ }\textbf {\bibinfo {volume}
  {3}},\ \bibinfo {pages} {L032064} (\bibinfo {year} {2021})}\BibitemShut
  {NoStop}%
\bibitem [{\citenamefont {Gaudin}(1967)}]{gaudin1967systeme}%
  \BibitemOpen
  \bibfield  {author} {\bibinfo {author} {\bibfnamefont {M.}~\bibnamefont
  {Gaudin}},\ }\href@noop {} {\bibfield  {journal} {\bibinfo  {journal}
  {Physics Letters A}\ }\textbf {\bibinfo {volume} {24}},\ \bibinfo {pages}
  {55} (\bibinfo {year} {1967})}\BibitemShut {NoStop}%
\bibitem [{\citenamefont {Yang}(1967)}]{yang1967some}%
  \BibitemOpen
  \bibfield  {author} {\bibinfo {author} {\bibfnamefont {C.~N.}\ \bibnamefont
  {Yang}},\ }\href {https://doi.org/10.1103/PhysRevLett.19.1312} {\bibfield
  {journal} {\bibinfo  {journal} {Phys. Rev. Lett.}\ }\textbf {\bibinfo
  {volume} {19}},\ \bibinfo {pages} {1312} (\bibinfo {year}
  {1967})}\BibitemShut {NoStop}%
\bibitem [{\citenamefont {Lieb}\ and\ \citenamefont
  {Liniger}(1963)}]{lieb1963exact}%
  \BibitemOpen
  \bibfield  {author} {\bibinfo {author} {\bibfnamefont {E.~H.}\ \bibnamefont
  {Lieb}}\ and\ \bibinfo {author} {\bibfnamefont {W.}~\bibnamefont {Liniger}},\
  }\href {https://doi.org/10.1103/PhysRev.130.1605} {\bibfield  {journal}
  {\bibinfo  {journal} {Phys. Rev.}\ }\textbf {\bibinfo {volume} {130}},\
  \bibinfo {pages} {1605} (\bibinfo {year} {1963})}\BibitemShut {NoStop}%
\bibitem [{\citenamefont {Moritz}\ \emph {et~al.}(2005)\citenamefont {Moritz},
  \citenamefont {St\"oferle}, \citenamefont {G\"unter}, \citenamefont
  {K\"ohl},\ and\ \citenamefont {Esslinger}}]{moritz2005confinement}%
  \BibitemOpen
  \bibfield  {author} {\bibinfo {author} {\bibfnamefont {H.}~\bibnamefont
  {Moritz}}, \bibinfo {author} {\bibfnamefont {T.}~\bibnamefont {St\"oferle}},
  \bibinfo {author} {\bibfnamefont {K.}~\bibnamefont {G\"unter}}, \bibinfo
  {author} {\bibfnamefont {M.}~\bibnamefont {K\"ohl}},\ and\ \bibinfo {author}
  {\bibfnamefont {T.}~\bibnamefont {Esslinger}},\ }\href
  {https://doi.org/10.1103/PhysRevLett.94.210401} {\bibfield  {journal}
  {\bibinfo  {journal} {Phys. Rev. Lett.}\ }\textbf {\bibinfo {volume} {94}},\
  \bibinfo {pages} {210401} (\bibinfo {year} {2005})}\BibitemShut {NoStop}%
\bibitem [{\citenamefont {Mora}\ \emph {et~al.}(2005)\citenamefont {Mora},
  \citenamefont {Komnik}, \citenamefont {Egger},\ and\ \citenamefont
  {Gogolin}}]{mora2005four}%
  \BibitemOpen
  \bibfield  {author} {\bibinfo {author} {\bibfnamefont {C.}~\bibnamefont
  {Mora}}, \bibinfo {author} {\bibfnamefont {A.}~\bibnamefont {Komnik}},
  \bibinfo {author} {\bibfnamefont {R.}~\bibnamefont {Egger}},\ and\ \bibinfo
  {author} {\bibfnamefont {A.~O.}\ \bibnamefont {Gogolin}},\ }\href
  {https://doi.org/10.1103/PhysRevLett.95.080403} {\bibfield  {journal}
  {\bibinfo  {journal} {Phys. Rev. Lett.}\ }\textbf {\bibinfo {volume} {95}},\
  \bibinfo {pages} {080403} (\bibinfo {year} {2005})}\BibitemShut {NoStop}%
\bibitem [{\citenamefont {Kinoshita}\ \emph {et~al.}(2006)\citenamefont
  {Kinoshita}, \citenamefont {Wenger},\ and\ \citenamefont
  {Weiss}}]{kinoshita2006quantum}%
  \BibitemOpen
  \bibfield  {author} {\bibinfo {author} {\bibfnamefont {T.}~\bibnamefont
  {Kinoshita}}, \bibinfo {author} {\bibfnamefont {T.}~\bibnamefont {Wenger}},\
  and\ \bibinfo {author} {\bibfnamefont {D.~S.}\ \bibnamefont {Weiss}},\ }\href
  {https://doi.org/10.1038/nature04693} {\bibfield  {journal} {\bibinfo
  {journal} {Nature}\ }\textbf {\bibinfo {volume} {440}},\ \bibinfo {pages}
  {900} (\bibinfo {year} {2006})}\BibitemShut {NoStop}%
\bibitem [{\citenamefont {{Langen}}\ \emph {et~al.}(2015)\citenamefont
  {{Langen}}, \citenamefont {{Erne}}, \citenamefont {{Geiger}}, \citenamefont
  {{Rauer}}, \citenamefont {{Schweigler}}, \citenamefont {{Kuhnert}},
  \citenamefont {{Rohringer}}, \citenamefont {{Mazets}}, \citenamefont
  {{Gasenzer}},\ and\ \citenamefont {{Schmiedmayer}}}]{langen2015experimental}%
  \BibitemOpen
  \bibfield  {author} {\bibinfo {author} {\bibfnamefont {T.}~\bibnamefont
  {{Langen}}}, \bibinfo {author} {\bibfnamefont {S.}~\bibnamefont {{Erne}}},
  \bibinfo {author} {\bibfnamefont {R.}~\bibnamefont {{Geiger}}}, \bibinfo
  {author} {\bibfnamefont {B.}~\bibnamefont {{Rauer}}}, \bibinfo {author}
  {\bibfnamefont {T.}~\bibnamefont {{Schweigler}}}, \bibinfo {author}
  {\bibfnamefont {M.}~\bibnamefont {{Kuhnert}}}, \bibinfo {author}
  {\bibfnamefont {W.}~\bibnamefont {{Rohringer}}}, \bibinfo {author}
  {\bibfnamefont {I.~E.}\ \bibnamefont {{Mazets}}}, \bibinfo {author}
  {\bibfnamefont {T.}~\bibnamefont {{Gasenzer}}},\ and\ \bibinfo {author}
  {\bibfnamefont {J.}~\bibnamefont {{Schmiedmayer}}},\ }\href
  {https://doi.org/10.1126/science.1257026} {\bibfield  {journal} {\bibinfo
  {journal} {Science}\ }\textbf {\bibinfo {volume} {348}},\ \bibinfo {pages}
  {207} (\bibinfo {year} {2015})}\BibitemShut {NoStop}%
\bibitem [{\citenamefont {Schemmer}\ \emph {et~al.}(2019)\citenamefont
  {Schemmer}, \citenamefont {Bouchoule}, \citenamefont {Doyon},\ and\
  \citenamefont {Dubail}}]{schemmer2019generalized}%
  \BibitemOpen
  \bibfield  {author} {\bibinfo {author} {\bibfnamefont {M.}~\bibnamefont
  {Schemmer}}, \bibinfo {author} {\bibfnamefont {I.}~\bibnamefont {Bouchoule}},
  \bibinfo {author} {\bibfnamefont {B.}~\bibnamefont {Doyon}},\ and\ \bibinfo
  {author} {\bibfnamefont {J.}~\bibnamefont {Dubail}},\ }\href
  {https://doi.org/10.1103/PhysRevLett.122.090601} {\bibfield  {journal}
  {\bibinfo  {journal} {Phys. Rev. Lett.}\ }\textbf {\bibinfo {volume} {122}},\
  \bibinfo {pages} {090601} (\bibinfo {year} {2019})}\BibitemShut {NoStop}%
\bibitem [{\citenamefont {Bouchoule}\ and\ \citenamefont
  {Dubail}(2022)}]{bouchoule2022generalized}%
  \BibitemOpen
  \bibfield  {author} {\bibinfo {author} {\bibfnamefont {I.}~\bibnamefont
  {Bouchoule}}\ and\ \bibinfo {author} {\bibfnamefont {J.}~\bibnamefont
  {Dubail}},\ }\href {https://doi.org/10.1088/1742-5468/ac3659} {\bibfield
  {journal} {\bibinfo  {journal} {J. Stat. Mech. Theory Exp.}\ }\textbf
  {\bibinfo {volume} {2022}},\ \bibinfo {pages} {014003} (\bibinfo {year}
  {2022})}\BibitemShut {NoStop}%
\bibitem [{\citenamefont {{Malvania}}\ \emph {et~al.}(2021)\citenamefont
  {{Malvania}}, \citenamefont {{Zhang}}, \citenamefont {{Le}}, \citenamefont
  {{Dubail}}, \citenamefont {{Rigol}},\ and\ \citenamefont
  {{Weiss}}}]{malvania2021generalized}%
  \BibitemOpen
  \bibfield  {author} {\bibinfo {author} {\bibfnamefont {N.}~\bibnamefont
  {{Malvania}}}, \bibinfo {author} {\bibfnamefont {Y.}~\bibnamefont {{Zhang}}},
  \bibinfo {author} {\bibfnamefont {Y.}~\bibnamefont {{Le}}}, \bibinfo {author}
  {\bibfnamefont {J.}~\bibnamefont {{Dubail}}}, \bibinfo {author}
  {\bibfnamefont {M.}~\bibnamefont {{Rigol}}},\ and\ \bibinfo {author}
  {\bibfnamefont {D.~S.}\ \bibnamefont {{Weiss}}},\ }\href
  {https://doi.org/10.1126/science.abf0147} {\bibfield  {journal} {\bibinfo
  {journal} {Science}\ }\textbf {\bibinfo {volume} {373}},\ \bibinfo {pages}
  {1129} (\bibinfo {year} {2021})}\BibitemShut {NoStop}%
\bibitem [{\citenamefont {{Schneider}}\ \emph {et~al.}(2012)\citenamefont
  {{Schneider}}, \citenamefont {{Hackerm{\"u}ller}}, \citenamefont
  {{Ronzheimer}}, \citenamefont {{Will}}, \citenamefont {{Braun}},
  \citenamefont {{Best}}, \citenamefont {{Bloch}}, \citenamefont {{Demler}},
  \citenamefont {{Mandt}}, \citenamefont {{Rasch}},\ and\ \citenamefont
  {{Rosch}}}]{schneider2012fermionic}%
  \BibitemOpen
  \bibfield  {author} {\bibinfo {author} {\bibfnamefont {U.}~\bibnamefont
  {{Schneider}}}, \bibinfo {author} {\bibfnamefont {L.}~\bibnamefont
  {{Hackerm{\"u}ller}}}, \bibinfo {author} {\bibfnamefont {J.~P.}\ \bibnamefont
  {{Ronzheimer}}}, \bibinfo {author} {\bibfnamefont {S.}~\bibnamefont
  {{Will}}}, \bibinfo {author} {\bibfnamefont {S.}~\bibnamefont {{Braun}}},
  \bibinfo {author} {\bibfnamefont {T.}~\bibnamefont {{Best}}}, \bibinfo
  {author} {\bibfnamefont {I.}~\bibnamefont {{Bloch}}}, \bibinfo {author}
  {\bibfnamefont {E.}~\bibnamefont {{Demler}}}, \bibinfo {author}
  {\bibfnamefont {S.}~\bibnamefont {{Mandt}}}, \bibinfo {author} {\bibfnamefont
  {D.}~\bibnamefont {{Rasch}}},\ and\ \bibinfo {author} {\bibfnamefont
  {A.}~\bibnamefont {{Rosch}}},\ }\href {https://doi.org/10.1038/nphys2205}
  {\bibfield  {journal} {\bibinfo  {journal} {Nat. Phys.}\ }\textbf {\bibinfo
  {volume} {8}},\ \bibinfo {pages} {213} (\bibinfo {year} {2012})}\BibitemShut
  {NoStop}%
\bibitem [{\citenamefont {{Hackerm{\"u}ller}}\ \emph
  {et~al.}(2010)\citenamefont {{Hackerm{\"u}ller}}, \citenamefont
  {{Schneider}}, \citenamefont {{Moreno-Cardoner}}, \citenamefont {{Kitagawa}},
  \citenamefont {{Best}}, \citenamefont {{Will}}, \citenamefont {{Demler}},
  \citenamefont {{Altman}}, \citenamefont {{Bloch}},\ and\ \citenamefont
  {{Paredes}}}]{hackerman2010anomalous}%
  \BibitemOpen
  \bibfield  {author} {\bibinfo {author} {\bibfnamefont {L.}~\bibnamefont
  {{Hackerm{\"u}ller}}}, \bibinfo {author} {\bibfnamefont {U.}~\bibnamefont
  {{Schneider}}}, \bibinfo {author} {\bibfnamefont {M.}~\bibnamefont
  {{Moreno-Cardoner}}}, \bibinfo {author} {\bibfnamefont {T.}~\bibnamefont
  {{Kitagawa}}}, \bibinfo {author} {\bibfnamefont {T.}~\bibnamefont {{Best}}},
  \bibinfo {author} {\bibfnamefont {S.}~\bibnamefont {{Will}}}, \bibinfo
  {author} {\bibfnamefont {E.}~\bibnamefont {{Demler}}}, \bibinfo {author}
  {\bibfnamefont {E.}~\bibnamefont {{Altman}}}, \bibinfo {author}
  {\bibfnamefont {I.}~\bibnamefont {{Bloch}}},\ and\ \bibinfo {author}
  {\bibfnamefont {B.}~\bibnamefont {{Paredes}}},\ }\href
  {https://doi.org/10.1126/science.1184565} {\bibfield  {journal} {\bibinfo
  {journal} {Science}\ }\textbf {\bibinfo {volume} {327}},\ \bibinfo {pages}
  {1621} (\bibinfo {year} {2010})}\BibitemShut {NoStop}%
\bibitem [{\citenamefont {Polkovnikov}\ \emph {et~al.}(2011)\citenamefont
  {Polkovnikov}, \citenamefont {Sengupta}, \citenamefont {Silva},\ and\
  \citenamefont {Vengalattore}}]{PolkovnikovReview}%
  \BibitemOpen
  \bibfield  {author} {\bibinfo {author} {\bibfnamefont {A.}~\bibnamefont
  {Polkovnikov}}, \bibinfo {author} {\bibfnamefont {K.}~\bibnamefont
  {Sengupta}}, \bibinfo {author} {\bibfnamefont {A.}~\bibnamefont {Silva}},\
  and\ \bibinfo {author} {\bibfnamefont {M.}~\bibnamefont {Vengalattore}},\
  }\href {https://doi.org/10.1103/RevModPhys.83.863} {\bibfield  {journal}
  {\bibinfo  {journal} {Rev. Mod. Phys.}\ }\textbf {\bibinfo {volume} {83}},\
  \bibinfo {pages} {863} (\bibinfo {year} {2011})}\BibitemShut {NoStop}%
\bibitem [{\citenamefont {Calabrese}\ \emph {et~al.}(2016)\citenamefont
  {Calabrese}, \citenamefont {Essler},\ and\ \citenamefont
  {Mussardo}}]{calabrese2016introduction}%
  \BibitemOpen
  \bibfield  {author} {\bibinfo {author} {\bibfnamefont {P.}~\bibnamefont
  {Calabrese}}, \bibinfo {author} {\bibfnamefont {F.~H.~L.}\ \bibnamefont
  {Essler}},\ and\ \bibinfo {author} {\bibfnamefont {G.}~\bibnamefont
  {Mussardo}},\ }\href {https://doi.org/10.1088/1742-5468/2016/06/064001}
  {\bibfield  {journal} {\bibinfo  {journal} {J. Stat. Mech. Theory Exp.}\
  }\textbf {\bibinfo {volume} {2016}},\ \bibinfo {pages} {064001} (\bibinfo
  {year} {2016})}\BibitemShut {NoStop}%
\bibitem [{\citenamefont {Vidmar}\ and\ \citenamefont
  {Rigol}(2016)}]{VidmarRigol}%
  \BibitemOpen
  \bibfield  {author} {\bibinfo {author} {\bibfnamefont {L.}~\bibnamefont
  {Vidmar}}\ and\ \bibinfo {author} {\bibfnamefont {M.}~\bibnamefont {Rigol}},\
  }\href {https://doi.org/10.1088/1742-5468/2016/06/064007} {\bibfield
  {journal} {\bibinfo  {journal} {J. Stat. Mech. Theory Exp.}\ }\textbf
  {\bibinfo {volume} {2016}},\ \bibinfo {pages} {064007} (\bibinfo {year}
  {2016})}\BibitemShut {NoStop}%
\bibitem [{\citenamefont {Essler}\ and\ \citenamefont
  {Fagotti}(2016)}]{essler2016quench}%
  \BibitemOpen
  \bibfield  {author} {\bibinfo {author} {\bibfnamefont {F.~H.~L.}\
  \bibnamefont {Essler}}\ and\ \bibinfo {author} {\bibfnamefont
  {M.}~\bibnamefont {Fagotti}},\ }\href
  {https://doi.org/10.1088/1742-5468/2016/06/064002} {\bibfield  {journal}
  {\bibinfo  {journal} {J. Stat. Mech. Theory Exp.}\ }\textbf {\bibinfo
  {volume} {2016}},\ \bibinfo {pages} {064002} (\bibinfo {year}
  {2016})}\BibitemShut {NoStop}%
\bibitem [{\citenamefont {Doyon}(2020)}]{doyon2020lecture}%
  \BibitemOpen
  \bibfield  {author} {\bibinfo {author} {\bibfnamefont {B.}~\bibnamefont
  {Doyon}},\ }\href {https://doi.org/10.21468/SciPostPhysLectNotes.18}
  {\bibfield  {journal} {\bibinfo  {journal} {SciPost Phys. Lect. Notes}\ ,\
  \bibinfo {pages} {18}} (\bibinfo {year} {2020})}\BibitemShut {NoStop}%
\bibitem [{\citenamefont {Bastianello}\ \emph {et~al.}(2022)\citenamefont
  {Bastianello}, \citenamefont {Bertini}, \citenamefont {Doyon},\ and\
  \citenamefont {Vasseur}}]{bastianello2022introduction}%
  \BibitemOpen
  \bibfield  {author} {\bibinfo {author} {\bibfnamefont {A.}~\bibnamefont
  {Bastianello}}, \bibinfo {author} {\bibfnamefont {B.}~\bibnamefont
  {Bertini}}, \bibinfo {author} {\bibfnamefont {B.}~\bibnamefont {Doyon}},\
  and\ \bibinfo {author} {\bibfnamefont {R.}~\bibnamefont {Vasseur}},\ }\href
  {https://doi.org/10.1088/1742-5468/ac3e6a} {\bibfield  {journal} {\bibinfo
  {journal} {J. Stat. Mech. Theory Exp.}\ }\textbf {\bibinfo {volume} {2022}},\
  \bibinfo {pages} {014001} (\bibinfo {year} {2022})}\BibitemShut {NoStop}%
\bibitem [{\citenamefont {Alba}\ \emph {et~al.}(2021)\citenamefont {Alba},
  \citenamefont {Bertini}, \citenamefont {Fagotti}, \citenamefont {Piroli},\
  and\ \citenamefont {Ruggiero}}]{alba2021generalized}%
  \BibitemOpen
  \bibfield  {author} {\bibinfo {author} {\bibfnamefont {V.}~\bibnamefont
  {Alba}}, \bibinfo {author} {\bibfnamefont {B.}~\bibnamefont {Bertini}},
  \bibinfo {author} {\bibfnamefont {M.}~\bibnamefont {Fagotti}}, \bibinfo
  {author} {\bibfnamefont {L.}~\bibnamefont {Piroli}},\ and\ \bibinfo {author}
  {\bibfnamefont {P.}~\bibnamefont {Ruggiero}},\ }\href
  {https://doi.org/10.1088/1742-5468/ac257d} {\bibfield  {journal} {\bibinfo
  {journal} {J. Stat. Mech. Theory Exp.}\ }\textbf {\bibinfo {volume} {2021}},\
  \bibinfo {pages} {114004} (\bibinfo {year} {2021})}\BibitemShut {NoStop}%
\bibitem [{\citenamefont {Calabrese}\ and\ \citenamefont
  {Cardy}(2005)}]{CalabreseCardy}%
  \BibitemOpen
  \bibfield  {author} {\bibinfo {author} {\bibfnamefont {P.}~\bibnamefont
  {Calabrese}}\ and\ \bibinfo {author} {\bibfnamefont {J.}~\bibnamefont
  {Cardy}},\ }\href {https://doi.org/10.1088/1742-5468/2005/04/p04010}
  {\bibfield  {journal} {\bibinfo  {journal} {J. Stat. Mech. Theory Exp.}\
  }\textbf {\bibinfo {volume} {2005}},\ \bibinfo {pages} {P04010} (\bibinfo
  {year} {2005})}\BibitemShut {NoStop}%
\bibitem [{\citenamefont {Calabrese}\ and\ \citenamefont
  {Cardy}(2016)}]{calabrese2016quantum}%
  \BibitemOpen
  \bibfield  {author} {\bibinfo {author} {\bibfnamefont {P.}~\bibnamefont
  {Calabrese}}\ and\ \bibinfo {author} {\bibfnamefont {J.}~\bibnamefont
  {Cardy}},\ }\href {https://doi.org/10.1088/1742-5468/2016/06/064003}
  {\bibfield  {journal} {\bibinfo  {journal} {J. Stat. Mech. Theory Exp.}\
  }\textbf {\bibinfo {volume} {2016}},\ \bibinfo {pages} {064003} (\bibinfo
  {year} {2016})}\BibitemShut {NoStop}%
\bibitem [{\citenamefont {{Alba}}\ and\ \citenamefont
  {{Calabrese}}(2017)}]{AlbaCalabrese1}%
  \BibitemOpen
  \bibfield  {author} {\bibinfo {author} {\bibfnamefont {V.}~\bibnamefont
  {{Alba}}}\ and\ \bibinfo {author} {\bibfnamefont {P.}~\bibnamefont
  {{Calabrese}}},\ }\href {https://doi.org/10.1073/pnas.1703516114} {\bibfield
  {journal} {\bibinfo  {journal} {PNAS}\ }\textbf {\bibinfo {volume} {114}},\
  \bibinfo {pages} {7947} (\bibinfo {year} {2017})}\BibitemShut {NoStop}%
\bibitem [{\citenamefont {Calabrese}(2020)}]{Calabrese2020}%
  \BibitemOpen
  \bibfield  {author} {\bibinfo {author} {\bibfnamefont {P.}~\bibnamefont
  {Calabrese}},\ }\href {https://doi.org/10.21468/SciPostPhysLectNotes.20}
  {\bibfield  {journal} {\bibinfo  {journal} {SciPost Phys. Lect. Notes}\ ,\
  \bibinfo {pages} {20}} (\bibinfo {year} {2020})}\BibitemShut {NoStop}%
\bibitem [{\citenamefont {Bertini}\ \emph {et~al.}(2022)\citenamefont
  {Bertini}, \citenamefont {Klobas}, \citenamefont {Alba}, \citenamefont
  {Lagnese},\ and\ \citenamefont {Calabrese}}]{bertini2022growth}%
  \BibitemOpen
  \bibfield  {author} {\bibinfo {author} {\bibfnamefont {B.}~\bibnamefont
  {Bertini}}, \bibinfo {author} {\bibfnamefont {K.}~\bibnamefont {Klobas}},
  \bibinfo {author} {\bibfnamefont {V.}~\bibnamefont {Alba}}, \bibinfo {author}
  {\bibfnamefont {G.}~\bibnamefont {Lagnese}},\ and\ \bibinfo {author}
  {\bibfnamefont {P.}~\bibnamefont {Calabrese}},\ }\Eprint
  {https://arxiv.org/abs/arXiv:2203.17264} {arXiv:2203.17264}  (\bibinfo {year}
  {2022})\BibitemShut {NoStop}%
\bibitem [{\citenamefont {Nahum}\ \emph {et~al.}(2017)\citenamefont {Nahum},
  \citenamefont {Ruhman}, \citenamefont {Vijay},\ and\ \citenamefont
  {Haah}}]{nahum2017quantum}%
  \BibitemOpen
  \bibfield  {author} {\bibinfo {author} {\bibfnamefont {A.}~\bibnamefont
  {Nahum}}, \bibinfo {author} {\bibfnamefont {J.}~\bibnamefont {Ruhman}},
  \bibinfo {author} {\bibfnamefont {S.}~\bibnamefont {Vijay}},\ and\ \bibinfo
  {author} {\bibfnamefont {J.}~\bibnamefont {Haah}},\ }\href
  {https://doi.org/10.1103/PhysRevX.7.031016} {\bibfield  {journal} {\bibinfo
  {journal} {Phys. Rev. X}\ }\textbf {\bibinfo {volume} {7}},\ \bibinfo {pages}
  {031016} (\bibinfo {year} {2017})}\BibitemShut {NoStop}%
\bibitem [{\citenamefont {Zhou}\ and\ \citenamefont
  {Nahum}(2020)}]{zhou2020entanglement}%
  \BibitemOpen
  \bibfield  {author} {\bibinfo {author} {\bibfnamefont {T.}~\bibnamefont
  {Zhou}}\ and\ \bibinfo {author} {\bibfnamefont {A.}~\bibnamefont {Nahum}},\
  }\href {https://doi.org/10.1103/PhysRevX.10.031066} {\bibfield  {journal}
  {\bibinfo  {journal} {Phys. Rev. X}\ }\textbf {\bibinfo {volume} {10}},\
  \bibinfo {pages} {031066} (\bibinfo {year} {2020})}\BibitemShut {NoStop}%
\bibitem [{\citenamefont {Yuzbashyan}\ \emph {et~al.}(2015)\citenamefont
  {Yuzbashyan}, \citenamefont {Dzero}, \citenamefont {Gurarie},\ and\
  \citenamefont {Foster}}]{yuzbashyan2015quantum}%
  \BibitemOpen
  \bibfield  {author} {\bibinfo {author} {\bibfnamefont {E.~A.}\ \bibnamefont
  {Yuzbashyan}}, \bibinfo {author} {\bibfnamefont {M.}~\bibnamefont {Dzero}},
  \bibinfo {author} {\bibfnamefont {V.}~\bibnamefont {Gurarie}},\ and\ \bibinfo
  {author} {\bibfnamefont {M.~S.}\ \bibnamefont {Foster}},\ }\href
  {https://doi.org/10.1103/PhysRevA.91.033628} {\bibfield  {journal} {\bibinfo
  {journal} {Phys. Rev. A}\ }\textbf {\bibinfo {volume} {91}},\ \bibinfo
  {pages} {033628} (\bibinfo {year} {2015})}\BibitemShut {NoStop}%
\bibitem [{\citenamefont {Kelly}\ \emph {et~al.}(2022)\citenamefont {Kelly},
  \citenamefont {Thompson}, \citenamefont {Rey},\ and\ \citenamefont
  {Marino}}]{kelly2022resonant}%
  \BibitemOpen
  \bibfield  {author} {\bibinfo {author} {\bibfnamefont {S.~P.}\ \bibnamefont
  {Kelly}}, \bibinfo {author} {\bibfnamefont {J.~K.}\ \bibnamefont {Thompson}},
  \bibinfo {author} {\bibfnamefont {A.~M.}\ \bibnamefont {Rey}},\ and\ \bibinfo
  {author} {\bibfnamefont {J.}~\bibnamefont {Marino}}\ }\href
  {https://doi.org/10.48550/ARXIV.2202.05851} {10.48550/ARXIV.2202.05851}
  (\bibinfo {year} {2022})\BibitemShut {NoStop}%
\bibitem [{\citenamefont {{Takahashi}}(1999)}]{Takahashi}%
  \BibitemOpen
  \bibfield  {author} {\bibinfo {author} {\bibfnamefont {M.}~\bibnamefont
  {{Takahashi}}},\ }\href@noop {} {\emph {\bibinfo {title} {Thermodynamics of
  One-Dimensional Solvable Models, by Minoru Takahashi, Cambridge, UK:
  Cambridge University Press, 1999}}}\ (\bibinfo {year} {1999})\BibitemShut
  {NoStop}%
\bibitem [{Note1()}]{Note1}%
  \BibitemOpen
  \bibinfo {note} {See supplementary material that contains (i) A brief review
  of the Bethe Ansatz solution of the Gaudin-Yang model; (ii) An explicit
  calculation of the overlap between the Bethe states and the BEC
  state~\protect \textup {\hbox {\mathsurround \z@ \protect \normalfont
  (\ignorespaces \ref {eq:BEC}\unskip \@@italiccorr )}}; (iii) The explicit
  form of the quench action for quenches from the BEC state~\protect \textup
  {\hbox {\mathsurround \z@ \protect \normalfont (\ignorespaces \ref
  {eq:BEC}\unskip \@@italiccorr )}}; A derivation of $g_2(\infty )$ using
  Feynman-Hellmann Theorem.}\BibitemShut {Stop}%
\bibitem [{\citenamefont {Caux}\ and\ \citenamefont
  {Essler}(2013)}]{CauxEssler}%
  \BibitemOpen
  \bibfield  {author} {\bibinfo {author} {\bibfnamefont {J.-S.}\ \bibnamefont
  {Caux}}\ and\ \bibinfo {author} {\bibfnamefont {F.~H.~L.}\ \bibnamefont
  {Essler}},\ }\href {https://doi.org/10.1103/PhysRevLett.110.257203}
  {\bibfield  {journal} {\bibinfo  {journal} {Phys. Rev. Lett.}\ }\textbf
  {\bibinfo {volume} {110}},\ \bibinfo {pages} {257203} (\bibinfo {year}
  {2013})}\BibitemShut {NoStop}%
\bibitem [{\citenamefont {Caux}(2016)}]{Caux}%
  \BibitemOpen
  \bibfield  {author} {\bibinfo {author} {\bibfnamefont {J.-S.}\ \bibnamefont
  {Caux}},\ }\href {https://doi.org/10.1088/1742-5468/2016/06/064006}
  {\bibfield  {journal} {\bibinfo  {journal} {J. Stat. Mech. Theory Exp.}\
  }\textbf {\bibinfo {volume} {2016}},\ \bibinfo {pages} {064006} (\bibinfo
  {year} {2016})}\BibitemShut {NoStop}%
\bibitem [{Note2()}]{Note2}%
  \BibitemOpen
  \bibinfo {note} {The fact that the description in terms of strings survives
  in the thermodynamic limit is an \protect \emph {assumption}, called string
  hypothesis, see, e.g,~\cite {Takahashi}.}\BibitemShut {Stop}%
\bibitem [{\citenamefont {{Piroli}}\ \emph {et~al.}(2017)\citenamefont
  {{Piroli}}, \citenamefont {{Pozsgay}},\ and\ \citenamefont
  {{Vernier}}}]{PiroliPozsgayVernier}%
  \BibitemOpen
  \bibfield  {author} {\bibinfo {author} {\bibfnamefont {L.}~\bibnamefont
  {{Piroli}}}, \bibinfo {author} {\bibfnamefont {B.}~\bibnamefont
  {{Pozsgay}}},\ and\ \bibinfo {author} {\bibfnamefont {E.}~\bibnamefont
  {{Vernier}}},\ }\href {https://doi.org/10.1016/j.nuclphysb.2017.10.012}
  {\bibfield  {journal} {\bibinfo  {journal} {Nucl. Phys. B.}\ }\textbf
  {\bibinfo {volume} {925}},\ \bibinfo {pages} {362} (\bibinfo {year}
  {2017})}\BibitemShut {NoStop}%
\bibitem [{\citenamefont {De~Nardis}\ \emph {et~al.}(2014)\citenamefont
  {De~Nardis}, \citenamefont {Wouters}, \citenamefont {Brockmann},\ and\
  \citenamefont {Caux}}]{DeNardis}%
  \BibitemOpen
  \bibfield  {author} {\bibinfo {author} {\bibfnamefont {J.}~\bibnamefont
  {De~Nardis}}, \bibinfo {author} {\bibfnamefont {B.}~\bibnamefont {Wouters}},
  \bibinfo {author} {\bibfnamefont {M.}~\bibnamefont {Brockmann}},\ and\
  \bibinfo {author} {\bibfnamefont {J.-S.}\ \bibnamefont {Caux}},\ }\href
  {https://doi.org/10.1103/PhysRevA.89.033601} {\bibfield  {journal} {\bibinfo
  {journal} {Phys. Rev. A}\ }\textbf {\bibinfo {volume} {89}},\ \bibinfo
  {pages} {033601} (\bibinfo {year} {2014})}\BibitemShut {NoStop}%
\bibitem [{\citenamefont {Brockmann}\ \emph {et~al.}(2014)\citenamefont
  {Brockmann}, \citenamefont {Wouters}, \citenamefont {Fioretto}, \citenamefont
  {Nardis}, \citenamefont {Vlijm},\ and\ \citenamefont {Caux}}]{Brockmann}%
  \BibitemOpen
  \bibfield  {author} {\bibinfo {author} {\bibfnamefont {M.}~\bibnamefont
  {Brockmann}}, \bibinfo {author} {\bibfnamefont {B.}~\bibnamefont {Wouters}},
  \bibinfo {author} {\bibfnamefont {D.}~\bibnamefont {Fioretto}}, \bibinfo
  {author} {\bibfnamefont {J.~D.}\ \bibnamefont {Nardis}}, \bibinfo {author}
  {\bibfnamefont {R.}~\bibnamefont {Vlijm}},\ and\ \bibinfo {author}
  {\bibfnamefont {J.-S.}\ \bibnamefont {Caux}},\ }\href
  {https://doi.org/10.1088/1742-5468/2014/12/p12009} {\bibfield  {journal}
  {\bibinfo  {journal} {J. Stat. Mech. Theory Exp.}\ }\textbf {\bibinfo
  {volume} {2014}},\ \bibinfo {pages} {P12009} (\bibinfo {year}
  {2014})}\BibitemShut {NoStop}%
\bibitem [{\citenamefont {Wouters}\ \emph {et~al.}(2014)\citenamefont
  {Wouters}, \citenamefont {De~Nardis}, \citenamefont {Brockmann},
  \citenamefont {Fioretto}, \citenamefont {Rigol},\ and\ \citenamefont
  {Caux}}]{wouters2014quenching}%
  \BibitemOpen
  \bibfield  {author} {\bibinfo {author} {\bibfnamefont {B.}~\bibnamefont
  {Wouters}}, \bibinfo {author} {\bibfnamefont {J.}~\bibnamefont {De~Nardis}},
  \bibinfo {author} {\bibfnamefont {M.}~\bibnamefont {Brockmann}}, \bibinfo
  {author} {\bibfnamefont {D.}~\bibnamefont {Fioretto}}, \bibinfo {author}
  {\bibfnamefont {M.}~\bibnamefont {Rigol}},\ and\ \bibinfo {author}
  {\bibfnamefont {J.-S.}\ \bibnamefont {Caux}},\ }\href
  {https://doi.org/10.1103/PhysRevLett.113.117202} {\bibfield  {journal}
  {\bibinfo  {journal} {Phys. Rev. Lett.}\ }\textbf {\bibinfo {volume} {113}},\
  \bibinfo {pages} {117202} (\bibinfo {year} {2014})}\BibitemShut {NoStop}%
\bibitem [{\citenamefont {Pozsgay}\ \emph {et~al.}(2014)\citenamefont
  {Pozsgay}, \citenamefont {Mesty\'an}, \citenamefont {Werner}, \citenamefont
  {Kormos}, \citenamefont {Zar\'and},\ and\ \citenamefont
  {Tak\'acs}}]{pozsgay2014correlations}%
  \BibitemOpen
  \bibfield  {author} {\bibinfo {author} {\bibfnamefont {B.}~\bibnamefont
  {Pozsgay}}, \bibinfo {author} {\bibfnamefont {M.}~\bibnamefont {Mesty\'an}},
  \bibinfo {author} {\bibfnamefont {M.~A.}\ \bibnamefont {Werner}}, \bibinfo
  {author} {\bibfnamefont {M.}~\bibnamefont {Kormos}}, \bibinfo {author}
  {\bibfnamefont {G.}~\bibnamefont {Zar\'and}},\ and\ \bibinfo {author}
  {\bibfnamefont {G.}~\bibnamefont {Tak\'acs}},\ }\href
  {https://doi.org/10.1103/PhysRevLett.113.117203} {\bibfield  {journal}
  {\bibinfo  {journal} {Phys. Rev. Lett.}\ }\textbf {\bibinfo {volume} {113}},\
  \bibinfo {pages} {117203} (\bibinfo {year} {2014})}\BibitemShut {NoStop}%
\bibitem [{\citenamefont {Mesty{\'{a}}n}\ \emph {et~al.}(2015)\citenamefont
  {Mesty{\'{a}}n}, \citenamefont {Pozsgay}, \citenamefont {Tak{\'{a}}cs},\ and\
  \citenamefont {Werner}}]{MestyanPozsgayTakacsWerner}%
  \BibitemOpen
  \bibfield  {author} {\bibinfo {author} {\bibfnamefont {M.}~\bibnamefont
  {Mesty{\'{a}}n}}, \bibinfo {author} {\bibfnamefont {B.}~\bibnamefont
  {Pozsgay}}, \bibinfo {author} {\bibfnamefont {G.}~\bibnamefont
  {Tak{\'{a}}cs}},\ and\ \bibinfo {author} {\bibfnamefont {M.~A.}\ \bibnamefont
  {Werner}},\ }\href {https://doi.org/10.1088/1742-5468/2015/04/p04001}
  {\bibfield  {journal} {\bibinfo  {journal} {J. Stat. Mech. Theory Exp.}\
  }\textbf {\bibinfo {volume} {2015}},\ \bibinfo {pages} {P04001} (\bibinfo
  {year} {2015})}\BibitemShut {NoStop}%
\bibitem [{\citenamefont {Bertini}\ \emph {et~al.}(2016)\citenamefont
  {Bertini}, \citenamefont {Piroli},\ and\ \citenamefont
  {Calabrese}}]{bertini2016quantum}%
  \BibitemOpen
  \bibfield  {author} {\bibinfo {author} {\bibfnamefont {B.}~\bibnamefont
  {Bertini}}, \bibinfo {author} {\bibfnamefont {L.}~\bibnamefont {Piroli}},\
  and\ \bibinfo {author} {\bibfnamefont {P.}~\bibnamefont {Calabrese}},\ }\href
  {https://doi.org/10.1088/1742-5468/2016/06/063102} {\bibfield  {journal}
  {\bibinfo  {journal} {J. Stat. Mech. Theory Exp.}\ }\textbf {\bibinfo
  {volume} {2016}},\ \bibinfo {pages} {063102} (\bibinfo {year}
  {2016})}\BibitemShut {NoStop}%
\bibitem [{\citenamefont {Mesty\'an}\ \emph {et~al.}(2019)\citenamefont
  {Mesty\'an}, \citenamefont {Bertini}, \citenamefont {Piroli},\ and\
  \citenamefont {Calabrese}}]{MestyanBertiniPiroliCalabrese}%
  \BibitemOpen
  \bibfield  {author} {\bibinfo {author} {\bibfnamefont {M.}~\bibnamefont
  {Mesty\'an}}, \bibinfo {author} {\bibfnamefont {B.}~\bibnamefont {Bertini}},
  \bibinfo {author} {\bibfnamefont {L.}~\bibnamefont {Piroli}},\ and\ \bibinfo
  {author} {\bibfnamefont {P.}~\bibnamefont {Calabrese}},\ }\href
  {https://doi.org/10.1103/PhysRevB.99.014305} {\bibfield  {journal} {\bibinfo
  {journal} {Phys. Rev. B}\ }\textbf {\bibinfo {volume} {99}},\ \bibinfo
  {pages} {014305} (\bibinfo {year} {2019})}\BibitemShut {NoStop}%
\bibitem [{\citenamefont {Bertini}\ \emph {et~al.}(2014)\citenamefont
  {Bertini}, \citenamefont {Schuricht},\ and\ \citenamefont
  {Essler}}]{BertiniSchurichtEssler}%
  \BibitemOpen
  \bibfield  {author} {\bibinfo {author} {\bibfnamefont {B.}~\bibnamefont
  {Bertini}}, \bibinfo {author} {\bibfnamefont {D.}~\bibnamefont {Schuricht}},\
  and\ \bibinfo {author} {\bibfnamefont {F.~H.~L.}\ \bibnamefont {Essler}},\
  }\href {https://doi.org/10.1088/1742-5468/2014/10/p10035} {\bibfield
  {journal} {\bibinfo  {journal} {J. Stat. Mech. Theory Exp.}\ }\textbf
  {\bibinfo {volume} {2014}},\ \bibinfo {pages} {P10035} (\bibinfo {year}
  {2014})}\BibitemShut {NoStop}%
\bibitem [{\citenamefont {Bertini}\ \emph {et~al.}(2017)\citenamefont
  {Bertini}, \citenamefont {Tartaglia},\ and\ \citenamefont
  {Calabrese}}]{BertiniTartagliaCalabrese}%
  \BibitemOpen
  \bibfield  {author} {\bibinfo {author} {\bibfnamefont {B.}~\bibnamefont
  {Bertini}}, \bibinfo {author} {\bibfnamefont {E.}~\bibnamefont {Tartaglia}},\
  and\ \bibinfo {author} {\bibfnamefont {P.}~\bibnamefont {Calabrese}},\ }\href
  {https://doi.org/10.1088/1742-5468/aa8c2c} {\bibfield  {journal} {\bibinfo
  {journal} {J. Stat. Mech. Theory Exp.}\ }\textbf {\bibinfo {volume} {2017}},\
  \bibinfo {pages} {103107} (\bibinfo {year} {2017})}\BibitemShut {NoStop}%
\bibitem [{\citenamefont {{Piroli}}\ \emph
  {et~al.}(2019{\natexlab{a}})\citenamefont {{Piroli}}, \citenamefont
  {{Vernier}}, \citenamefont {{Calabrese}},\ and\ \citenamefont
  {{Pozsgay}}}]{PiroliVernierCalabresePozsgay1}%
  \BibitemOpen
  \bibfield  {author} {\bibinfo {author} {\bibfnamefont {L.}~\bibnamefont
  {{Piroli}}}, \bibinfo {author} {\bibfnamefont {E.}~\bibnamefont {{Vernier}}},
  \bibinfo {author} {\bibfnamefont {P.}~\bibnamefont {{Calabrese}}},\ and\
  \bibinfo {author} {\bibfnamefont {B.}~\bibnamefont {{Pozsgay}}},\ }\href
  {https://doi.org/10.1088/1742-5468/ab1c51} {\bibfield  {journal} {\bibinfo
  {journal} {J. Stat. Mech. Theory Exp.}\ }\textbf {\bibinfo {volume} {6}},\
  \bibinfo {pages} {063103} (\bibinfo {year} {2019}{\natexlab{a}})}\BibitemShut
  {NoStop}%
\bibitem [{\citenamefont {{Piroli}}\ \emph
  {et~al.}(2019{\natexlab{b}})\citenamefont {{Piroli}}, \citenamefont
  {{Vernier}}, \citenamefont {{Calabrese}},\ and\ \citenamefont
  {{Pozsgay}}}]{PiroliVernierCalabresePozsgay2}%
  \BibitemOpen
  \bibfield  {author} {\bibinfo {author} {\bibfnamefont {L.}~\bibnamefont
  {{Piroli}}}, \bibinfo {author} {\bibfnamefont {E.}~\bibnamefont {{Vernier}}},
  \bibinfo {author} {\bibfnamefont {P.}~\bibnamefont {{Calabrese}}},\ and\
  \bibinfo {author} {\bibfnamefont {B.}~\bibnamefont {{Pozsgay}}},\ }\href
  {https://doi.org/10.1088/1742-5468/ab1c52} {\bibfield  {journal} {\bibinfo
  {journal} {J. Stat. Mech. Theory Exp.}\ }\textbf {\bibinfo {volume} {6}},\
  \bibinfo {pages} {063104} (\bibinfo {year} {2019}{\natexlab{b}})}\BibitemShut
  {NoStop}%
\bibitem [{\citenamefont {Piroli}\ \emph
  {et~al.}(2016{\natexlab{a}})\citenamefont {Piroli}, \citenamefont
  {Calabrese},\ and\ \citenamefont {Essler}}]{piroli2016multiparticle}%
  \BibitemOpen
  \bibfield  {author} {\bibinfo {author} {\bibfnamefont {L.}~\bibnamefont
  {Piroli}}, \bibinfo {author} {\bibfnamefont {P.}~\bibnamefont {Calabrese}},\
  and\ \bibinfo {author} {\bibfnamefont {F.~H.~L.}\ \bibnamefont {Essler}},\
  }\href {https://doi.org/10.1103/PhysRevLett.116.070408} {\bibfield  {journal}
  {\bibinfo  {journal} {Phys. Rev. Lett.}\ }\textbf {\bibinfo {volume} {116}},\
  \bibinfo {pages} {070408} (\bibinfo {year} {2016}{\natexlab{a}})}\BibitemShut
  {NoStop}%
\bibitem [{\citenamefont {Piroli}\ \emph
  {et~al.}(2016{\natexlab{b}})\citenamefont {Piroli}, \citenamefont
  {Calabrese},\ and\ \citenamefont {Essler}}]{piroli2016quantum}%
  \BibitemOpen
  \bibfield  {author} {\bibinfo {author} {\bibfnamefont {L.}~\bibnamefont
  {Piroli}}, \bibinfo {author} {\bibfnamefont {P.}~\bibnamefont {Calabrese}},\
  and\ \bibinfo {author} {\bibfnamefont {F.~H.~L.}\ \bibnamefont {Essler}},\
  }\href {https://doi.org/10.21468/SciPostPhys.1.1.001} {\bibfield  {journal}
  {\bibinfo  {journal} {SciPost Phys.}\ }\textbf {\bibinfo {volume} {1}},\
  \bibinfo {pages} {001} (\bibinfo {year} {2016}{\natexlab{b}})}\BibitemShut
  {NoStop}%
\bibitem [{\citenamefont {Alba}\ and\ \citenamefont
  {Calabrese}(2016)}]{alba2016the}%
  \BibitemOpen
  \bibfield  {author} {\bibinfo {author} {\bibfnamefont {V.}~\bibnamefont
  {Alba}}\ and\ \bibinfo {author} {\bibfnamefont {P.}~\bibnamefont
  {Calabrese}},\ }\href {https://doi.org/10.1088/1742-5468/2016/04/043105}
  {\bibfield  {journal} {\bibinfo  {journal} {J. Stat. Mech. Theory Exp.}\
  }\textbf {\bibinfo {volume} {2016}},\ \bibinfo {pages} {043105} (\bibinfo
  {year} {2016})}\BibitemShut {NoStop}%
\bibitem [{\citenamefont {Piroli}\ \emph
  {et~al.}(2016{\natexlab{c}})\citenamefont {Piroli}, \citenamefont {Vernier},\
  and\ \citenamefont {Calabrese}}]{piroli2016exact}%
  \BibitemOpen
  \bibfield  {author} {\bibinfo {author} {\bibfnamefont {L.}~\bibnamefont
  {Piroli}}, \bibinfo {author} {\bibfnamefont {E.}~\bibnamefont {Vernier}},\
  and\ \bibinfo {author} {\bibfnamefont {P.}~\bibnamefont {Calabrese}},\ }\href
  {https://doi.org/10.1103/PhysRevB.94.054313} {\bibfield  {journal} {\bibinfo
  {journal} {Phys. Rev. B}\ }\textbf {\bibinfo {volume} {94}},\ \bibinfo
  {pages} {054313} (\bibinfo {year} {2016}{\natexlab{c}})}\BibitemShut
  {NoStop}%
\bibitem [{\citenamefont {Nardis}\ \emph {et~al.}(2015)\citenamefont {Nardis},
  \citenamefont {Piroli},\ and\ \citenamefont {Caux}}]{denardis2015relaxation}%
  \BibitemOpen
  \bibfield  {author} {\bibinfo {author} {\bibfnamefont {J.~D.}\ \bibnamefont
  {Nardis}}, \bibinfo {author} {\bibfnamefont {L.}~\bibnamefont {Piroli}},\
  and\ \bibinfo {author} {\bibfnamefont {J.-S.}\ \bibnamefont {Caux}},\ }\href
  {https://doi.org/10.1088/1751-8113/48/43/43ft01} {\bibfield  {journal}
  {\bibinfo  {journal} {J. Phys. A Math. Theor.}\ }\textbf {\bibinfo {volume}
  {48}},\ \bibinfo {pages} {43FT01} (\bibinfo {year} {2015})}\BibitemShut
  {NoStop}%
\bibitem [{\citenamefont {Mesty{\'{a}}n}\ \emph {et~al.}(2017)\citenamefont
  {Mesty{\'{a}}n}, \citenamefont {Bertini}, \citenamefont {Piroli},\ and\
  \citenamefont {Calabrese}}]{mestyan2017exact}%
  \BibitemOpen
  \bibfield  {author} {\bibinfo {author} {\bibfnamefont {M.}~\bibnamefont
  {Mesty{\'{a}}n}}, \bibinfo {author} {\bibfnamefont {B.}~\bibnamefont
  {Bertini}}, \bibinfo {author} {\bibfnamefont {L.}~\bibnamefont {Piroli}},\
  and\ \bibinfo {author} {\bibfnamefont {P.}~\bibnamefont {Calabrese}},\ }\href
  {https://doi.org/10.1088/1742-5468/aa7df0} {\bibfield  {journal} {\bibinfo
  {journal} {J. Stat. Mech. Theory Exp.}\ }\textbf {\bibinfo {volume} {2017}},\
  \bibinfo {pages} {083103} (\bibinfo {year} {2017})}\BibitemShut {NoStop}%
\bibitem [{\citenamefont {Rylands}\ and\ \citenamefont
  {Andrei}(2019)}]{rylands2019loschmidt}%
  \BibitemOpen
  \bibfield  {author} {\bibinfo {author} {\bibfnamefont {C.}~\bibnamefont
  {Rylands}}\ and\ \bibinfo {author} {\bibfnamefont {N.}~\bibnamefont
  {Andrei}},\ }\href {https://doi.org/10.1103/PhysRevB.99.085133} {\bibfield
  {journal} {\bibinfo  {journal} {Phys. Rev. B}\ }\textbf {\bibinfo {volume}
  {99}},\ \bibinfo {pages} {085133} (\bibinfo {year} {2019})}\BibitemShut
  {NoStop}%
\bibitem [{\citenamefont {{de Leeuw}}\ \emph {et~al.}(2018)\citenamefont {{de
  Leeuw}}, \citenamefont {{Kristjansen}},\ and\ \citenamefont
  {{Linardopoulos}}}]{deleeuw2018scalar}%
  \BibitemOpen
  \bibfield  {author} {\bibinfo {author} {\bibfnamefont {M.}~\bibnamefont {{de
  Leeuw}}}, \bibinfo {author} {\bibfnamefont {C.}~\bibnamefont
  {{Kristjansen}}},\ and\ \bibinfo {author} {\bibfnamefont {G.}~\bibnamefont
  {{Linardopoulos}}},\ }\href {https://doi.org/10.1016/j.physletb.2018.03.083}
  {\bibfield  {journal} {\bibinfo  {journal} {Physics Letters B}\ }\textbf
  {\bibinfo {volume} {781}},\ \bibinfo {pages} {238} (\bibinfo {year}
  {2018})},\ \Eprint {https://arxiv.org/abs/1802.01598} {arXiv:1802.01598
  [hep-th]} \BibitemShut {NoStop}%
\bibitem [{\citenamefont {{de Leeuw}}\ \emph {et~al.}(2020)\citenamefont {{de
  Leeuw}}, \citenamefont {{Gombor}}, \citenamefont {{Kristjansen}},
  \citenamefont {{Linardopoulos}},\ and\ \citenamefont
  {{Pozsgay}}}]{deleeuw2020spin}%
  \BibitemOpen
  \bibfield  {author} {\bibinfo {author} {\bibfnamefont {M.}~\bibnamefont {{de
  Leeuw}}}, \bibinfo {author} {\bibfnamefont {T.}~\bibnamefont {{Gombor}}},
  \bibinfo {author} {\bibfnamefont {C.}~\bibnamefont {{Kristjansen}}}, \bibinfo
  {author} {\bibfnamefont {G.}~\bibnamefont {{Linardopoulos}}},\ and\ \bibinfo
  {author} {\bibfnamefont {B.}~\bibnamefont {{Pozsgay}}},\ }\href
  {https://doi.org/10.1007/JHEP01(2020)176} {\bibfield  {journal} {\bibinfo
  {journal} {Journal of High Energy Physics}\ }\textbf {\bibinfo {volume}
  {2020}},\ \bibinfo {eid} {176} (\bibinfo {year} {2020})},\ \Eprint
  {https://arxiv.org/abs/1912.09338} {arXiv:1912.09338 [hep-th]} \BibitemShut
  {NoStop}%
\bibitem [{\citenamefont {{de Leeuw}}\ \emph {et~al.}(2017)\citenamefont {{de
  Leeuw}}, \citenamefont {{Kristjansen}},\ and\ \citenamefont
  {{Linardopoulos}}}]{deleeuw2016one}%
  \BibitemOpen
  \bibfield  {author} {\bibinfo {author} {\bibfnamefont {M.}~\bibnamefont {{de
  Leeuw}}}, \bibinfo {author} {\bibfnamefont {C.}~\bibnamefont
  {{Kristjansen}}},\ and\ \bibinfo {author} {\bibfnamefont {G.}~\bibnamefont
  {{Linardopoulos}}},\ }\href {https://doi.org/10.1088/1751-8121/aa714b}
  {\bibfield  {journal} {\bibinfo  {journal} {Journal of Physics A Mathematical
  General}\ }\textbf {\bibinfo {volume} {50}},\ \bibinfo {eid} {254001}
  (\bibinfo {year} {2017})},\ \Eprint {https://arxiv.org/abs/1612.06236}
  {arXiv:1612.06236 [hep-th]} \BibitemShut {NoStop}%
\bibitem [{\citenamefont {{Kristjansen}}\ \emph {et~al.}(2022)\citenamefont
  {{Kristjansen}}, \citenamefont {{Vu}},\ and\ \citenamefont
  {{Zarembo}}}]{kristjansen2022integrable}%
  \BibitemOpen
  \bibfield  {author} {\bibinfo {author} {\bibfnamefont {C.}~\bibnamefont
  {{Kristjansen}}}, \bibinfo {author} {\bibfnamefont {D.-L.}\ \bibnamefont
  {{Vu}}},\ and\ \bibinfo {author} {\bibfnamefont {K.}~\bibnamefont
  {{Zarembo}}},\ }\href {https://doi.org/10.1007/JHEP02(2022)070} {\bibfield
  {journal} {\bibinfo  {journal} {Journal of High Energy Physics}\ }\textbf
  {\bibinfo {volume} {2022}},\ \bibinfo {eid} {70} (\bibinfo {year} {2022})},\
  \Eprint {https://arxiv.org/abs/2112.10438} {arXiv:2112.10438 [hep-th]}
  \BibitemShut {NoStop}%
\bibitem [{\citenamefont {Rylands}\ \emph {et~al.}(2022)\citenamefont
  {Rylands}, \citenamefont {Bertini},\ and\ \citenamefont
  {Calabrese}}]{rylands2022integrable}%
  \BibitemOpen
  \bibfield  {author} {\bibinfo {author} {\bibfnamefont {C.}~\bibnamefont
  {Rylands}}, \bibinfo {author} {\bibfnamefont {B.}~\bibnamefont {Bertini}},\
  and\ \bibinfo {author} {\bibfnamefont {P.}~\bibnamefont {Calabrese}},\ }\href
  {https://doi.org/10.1088/1742-5468/ac98be} {\bibfield  {journal} {\bibinfo
  {journal} {J. Stat. Mech. Theory Exp.}\ }\textbf {\bibinfo {volume} {2022}},\
  \bibinfo {pages} {103103} (\bibinfo {year} {2022})}\BibitemShut {NoStop}%
\bibitem [{\citenamefont {{Essler}}\ \emph {et~al.}(2005)\citenamefont
  {{Essler}}, \citenamefont {{Frahm}}, \citenamefont {{G{\"o}hmann}},
  \citenamefont {{Kl{\"u}mper}},\ and\ \citenamefont
  {{Korepin}}}]{EsslerFrahmGohmannKlumper}%
  \BibitemOpen
  \bibfield  {author} {\bibinfo {author} {\bibfnamefont {F.~H.~L.}\
  \bibnamefont {{Essler}}}, \bibinfo {author} {\bibfnamefont {H.}~\bibnamefont
  {{Frahm}}}, \bibinfo {author} {\bibfnamefont {F.}~\bibnamefont
  {{G{\"o}hmann}}}, \bibinfo {author} {\bibfnamefont {A.}~\bibnamefont
  {{Kl{\"u}mper}}},\ and\ \bibinfo {author} {\bibfnamefont {V.~E.}\
  \bibnamefont {{Korepin}}},\ }\href@noop {} {\emph {\bibinfo {title} {{The
  One-Dimensional Hubbard Model}}}}\ (\bibinfo {year} {2005})\BibitemShut
  {NoStop}%
\bibitem [{\citenamefont {Brockmann}(2014)}]{brockmann2014overlaps}%
  \BibitemOpen
  \bibfield  {author} {\bibinfo {author} {\bibfnamefont {M.}~\bibnamefont
  {Brockmann}},\ }\href {https://doi.org/10.1088/1742-5468/2014/05/p05006}
  {\bibfield  {journal} {\bibinfo  {journal} {J. Stat. Mech. Theory Exp.}\
  }\textbf {\bibinfo {volume} {2014}},\ \bibinfo {pages} {P05006} (\bibinfo
  {year} {2014})}\BibitemShut {NoStop}%
\bibitem [{\citenamefont {Mesty{\'{a}}n}\ and\ \citenamefont
  {Pozsgay}(2014)}]{mestyan2014short}%
  \BibitemOpen
  \bibfield  {author} {\bibinfo {author} {\bibfnamefont {M.}~\bibnamefont
  {Mesty{\'{a}}n}}\ and\ \bibinfo {author} {\bibfnamefont {B.}~\bibnamefont
  {Pozsgay}},\ }\href {https://doi.org/10.1088/1742-5468/2014/09/p09020}
  {\bibfield  {journal} {\bibinfo  {journal} {J. Stat. Mech. Theory Exp.}\
  }\textbf {\bibinfo {volume} {2014}},\ \bibinfo {pages} {P09020} (\bibinfo
  {year} {2014})}\BibitemShut {NoStop}%
\bibitem [{\citenamefont {Negro}\ and\ \citenamefont
  {Smirnov}(2013)}]{negro2013on}%
  \BibitemOpen
  \bibfield  {author} {\bibinfo {author} {\bibfnamefont {S.}~\bibnamefont
  {Negro}}\ and\ \bibinfo {author} {\bibfnamefont {F.}~\bibnamefont
  {Smirnov}},\ }\href
  {https://doi.org/https://doi.org/10.1016/j.nuclphysb.2013.06.023} {\bibfield
  {journal} {\bibinfo  {journal} {Nucl. Phys. B.}\ }\textbf {\bibinfo {volume}
  {875}},\ \bibinfo {pages} {166} (\bibinfo {year} {2013})}\BibitemShut
  {NoStop}%
\bibitem [{\citenamefont {Negro}(2014)}]{negro2014on}%
  \BibitemOpen
  \bibfield  {author} {\bibinfo {author} {\bibfnamefont {S.}~\bibnamefont
  {Negro}},\ }\href {https://doi.org/10.1142/S0217751X14501115} {\bibfield
  {journal} {\bibinfo  {journal} {Int. J. Mod. Phys. A}\ }\textbf {\bibinfo
  {volume} {29}},\ \bibinfo {pages} {1450111} (\bibinfo {year}
  {2014})}\BibitemShut {NoStop}%
\bibitem [{\citenamefont {Pozsgay}(2011)}]{pozsgay2011local}%
  \BibitemOpen
  \bibfield  {author} {\bibinfo {author} {\bibfnamefont {B.}~\bibnamefont
  {Pozsgay}},\ }\href {https://doi.org/10.1088/1742-5468/2011/11/p11017}
  {\bibfield  {journal} {\bibinfo  {journal} {J. Stat. Mech. Theory Exp.}\
  }\textbf {\bibinfo {volume} {2011}},\ \bibinfo {pages} {P11017} (\bibinfo
  {year} {2011})}\BibitemShut {NoStop}%
\bibitem [{\citenamefont {Bastianello}\ \emph {et~al.}(2018)\citenamefont
  {Bastianello}, \citenamefont {Piroli},\ and\ \citenamefont
  {Calabrese}}]{bastianello2018exact}%
  \BibitemOpen
  \bibfield  {author} {\bibinfo {author} {\bibfnamefont {A.}~\bibnamefont
  {Bastianello}}, \bibinfo {author} {\bibfnamefont {L.}~\bibnamefont
  {Piroli}},\ and\ \bibinfo {author} {\bibfnamefont {P.}~\bibnamefont
  {Calabrese}},\ }\href {https://doi.org/10.1103/PhysRevLett.120.190601}
  {\bibfield  {journal} {\bibinfo  {journal} {Phys. Rev. Lett.}\ }\textbf
  {\bibinfo {volume} {120}},\ \bibinfo {pages} {190601} (\bibinfo {year}
  {2018})}\BibitemShut {NoStop}%
\bibitem [{\citenamefont {Bastianello}\ and\ \citenamefont
  {Piroli}(2018)}]{bastianello2018from}%
  \BibitemOpen
  \bibfield  {author} {\bibinfo {author} {\bibfnamefont {A.}~\bibnamefont
  {Bastianello}}\ and\ \bibinfo {author} {\bibfnamefont {L.}~\bibnamefont
  {Piroli}},\ }\href {https://doi.org/10.1088/1742-5468/aaeb48} {\bibfield
  {journal} {\bibinfo  {journal} {J. Stat. Mech. Theory Exp.}\ }\textbf
  {\bibinfo {volume} {2018}},\ \bibinfo {pages} {113104} (\bibinfo {year}
  {2018})}\BibitemShut {NoStop}%
\bibitem [{\citenamefont {Tan}(2008{\natexlab{a}})}]{tan2008energetics}%
  \BibitemOpen
  \bibfield  {author} {\bibinfo {author} {\bibfnamefont {S.}~\bibnamefont
  {Tan}},\ }\href {https://doi.org/10.1016/j.aop.2008.03.004} {\bibfield
  {journal} {\bibinfo  {journal} {Annals of Physics}\ }\textbf {\bibinfo
  {volume} {323}},\ \bibinfo {pages} {2952} (\bibinfo {year}
  {2008}{\natexlab{a}})}\BibitemShut {NoStop}%
\bibitem [{\citenamefont {Tan}(2008{\natexlab{b}})}]{tan2008large}%
  \BibitemOpen
  \bibfield  {author} {\bibinfo {author} {\bibfnamefont {S.}~\bibnamefont
  {Tan}},\ }\href {https://doi.org/10.1016/j.aop.2008.03.005} {\bibfield
  {journal} {\bibinfo  {journal} {Annals of Physics}\ }\textbf {\bibinfo
  {volume} {323}},\ \bibinfo {pages} {2971} (\bibinfo {year}
  {2008}{\natexlab{b}})}\BibitemShut {NoStop}%
\bibitem [{\citenamefont {Tan}(2008{\natexlab{c}})}]{tan2008generalized}%
  \BibitemOpen
  \bibfield  {author} {\bibinfo {author} {\bibfnamefont {S.}~\bibnamefont
  {Tan}},\ }\href {https://doi.org/10.1016/j.aop.2008.03.003} {\bibfield
  {journal} {\bibinfo  {journal} {Annals of Physics}\ }\textbf {\bibinfo
  {volume} {323}},\ \bibinfo {pages} {2987} (\bibinfo {year}
  {2008}{\natexlab{c}})}\BibitemShut {NoStop}%
\bibitem [{\citenamefont {Barth}\ and\ \citenamefont
  {Zwerger}(2011)}]{barth2011tan}%
  \BibitemOpen
  \bibfield  {author} {\bibinfo {author} {\bibfnamefont {M.}~\bibnamefont
  {Barth}}\ and\ \bibinfo {author} {\bibfnamefont {W.}~\bibnamefont
  {Zwerger}},\ }\href {https://doi.org/10.1016/j.aop.2011.05.010} {\bibfield
  {journal} {\bibinfo  {journal} {Annals of Physics}\ }\textbf {\bibinfo
  {volume} {326}},\ \bibinfo {pages} {2544} (\bibinfo {year}
  {2011})}\BibitemShut {NoStop}%
\bibitem [{\citenamefont {Decamp}\ \emph {et~al.}(2016)\citenamefont {Decamp},
  \citenamefont {J\"unemann}, \citenamefont {Albert}, \citenamefont {Rizzi},
  \citenamefont {Minguzzi},\ and\ \citenamefont {Vignolo}}]{decamp2016high}%
  \BibitemOpen
  \bibfield  {author} {\bibinfo {author} {\bibfnamefont {J.}~\bibnamefont
  {Decamp}}, \bibinfo {author} {\bibfnamefont {J.}~\bibnamefont {J\"unemann}},
  \bibinfo {author} {\bibfnamefont {M.}~\bibnamefont {Albert}}, \bibinfo
  {author} {\bibfnamefont {M.}~\bibnamefont {Rizzi}}, \bibinfo {author}
  {\bibfnamefont {A.}~\bibnamefont {Minguzzi}},\ and\ \bibinfo {author}
  {\bibfnamefont {P.}~\bibnamefont {Vignolo}},\ }\href
  {https://doi.org/10.1103/PhysRevA.94.053614} {\bibfield  {journal} {\bibinfo
  {journal} {Phys. Rev. A}\ }\textbf {\bibinfo {volume} {94}},\ \bibinfo
  {pages} {053614} (\bibinfo {year} {2016})}\BibitemShut {NoStop}%
\bibitem [{\citenamefont {Vignolo}\ and\ \citenamefont
  {Minguzzi}(2013)}]{vignolo2013universal}%
  \BibitemOpen
  \bibfield  {author} {\bibinfo {author} {\bibfnamefont {P.}~\bibnamefont
  {Vignolo}}\ and\ \bibinfo {author} {\bibfnamefont {A.}~\bibnamefont
  {Minguzzi}},\ }\href {https://doi.org/10.1103/PhysRevLett.110.020403}
  {\bibfield  {journal} {\bibinfo  {journal} {Phys. Rev. Lett.}\ }\textbf
  {\bibinfo {volume} {110}},\ \bibinfo {pages} {020403} (\bibinfo {year}
  {2013})}\BibitemShut {NoStop}%
\bibitem [{\citenamefont {Bouchoule}\ and\ \citenamefont
  {Dubail}(2021)}]{bouchoule2021breakdown}%
  \BibitemOpen
  \bibfield  {author} {\bibinfo {author} {\bibfnamefont {I.}~\bibnamefont
  {Bouchoule}}\ and\ \bibinfo {author} {\bibfnamefont {J.}~\bibnamefont
  {Dubail}},\ }\href {https://doi.org/10.1103/PhysRevLett.126.160603}
  {\bibfield  {journal} {\bibinfo  {journal} {Phys. Rev. Lett.}\ }\textbf
  {\bibinfo {volume} {126}},\ \bibinfo {pages} {160603} (\bibinfo {year}
  {2021})}\BibitemShut {NoStop}%
\bibitem [{\citenamefont {Paintner}\ \emph {et~al.}(2019)\citenamefont
  {Paintner}, \citenamefont {Hoffmann}, \citenamefont {J\"ager}, \citenamefont
  {Limmer}, \citenamefont {Schoch}, \citenamefont {Deissler}, \citenamefont
  {Pini}, \citenamefont {Pieri}, \citenamefont {Calvanese~Strinati},
  \citenamefont {Chin},\ and\ \citenamefont
  {Hecker~Denschlag}}]{paintner2019pair}%
  \BibitemOpen
  \bibfield  {author} {\bibinfo {author} {\bibfnamefont {T.}~\bibnamefont
  {Paintner}}, \bibinfo {author} {\bibfnamefont {D.~K.}\ \bibnamefont
  {Hoffmann}}, \bibinfo {author} {\bibfnamefont {M.}~\bibnamefont {J\"ager}},
  \bibinfo {author} {\bibfnamefont {W.}~\bibnamefont {Limmer}}, \bibinfo
  {author} {\bibfnamefont {W.}~\bibnamefont {Schoch}}, \bibinfo {author}
  {\bibfnamefont {B.}~\bibnamefont {Deissler}}, \bibinfo {author}
  {\bibfnamefont {M.}~\bibnamefont {Pini}}, \bibinfo {author} {\bibfnamefont
  {P.}~\bibnamefont {Pieri}}, \bibinfo {author} {\bibfnamefont
  {G.}~\bibnamefont {Calvanese~Strinati}}, \bibinfo {author} {\bibfnamefont
  {C.}~\bibnamefont {Chin}},\ and\ \bibinfo {author} {\bibfnamefont
  {J.}~\bibnamefont {Hecker~Denschlag}},\ }\href
  {https://doi.org/10.1103/PhysRevA.99.053617} {\bibfield  {journal} {\bibinfo
  {journal} {Phys. Rev. A}\ }\textbf {\bibinfo {volume} {99}},\ \bibinfo
  {pages} {053617} (\bibinfo {year} {2019})}\BibitemShut {NoStop}%
\bibitem [{\citenamefont {Pini}\ \emph {et~al.}(2020)\citenamefont {Pini},
  \citenamefont {Pieri}, \citenamefont {JŠger}, \citenamefont {Denschlag},\
  and\ \citenamefont {Strinati}}]{pini2020pair}%
  \BibitemOpen
  \bibfield  {author} {\bibinfo {author} {\bibfnamefont {M.}~\bibnamefont
  {Pini}}, \bibinfo {author} {\bibfnamefont {P.}~\bibnamefont {Pieri}},
  \bibinfo {author} {\bibfnamefont {M.}~\bibnamefont {JŠger}}, \bibinfo
  {author} {\bibfnamefont {J.~H.}\ \bibnamefont {Denschlag}},\ and\ \bibinfo
  {author} {\bibfnamefont {G.~C.}\ \bibnamefont {Strinati}},\ }\href
  {https://doi.org/10.1088/1367-2630/ab9ee3} {\bibfield  {journal} {\bibinfo
  {journal} {New Journal of Physics}\ }\textbf {\bibinfo {volume} {22}},\
  \bibinfo {pages} {083008} (\bibinfo {year} {2020})}\BibitemShut {NoStop}%
\bibitem [{\citenamefont {Cataldini}\ \emph {et~al.}(2021)\citenamefont
  {Cataldini}, \citenamefont {M{\o}ller}, \citenamefont {Tajik}, \citenamefont
  {Sabino}, \citenamefont {Schweigler}, \citenamefont {Ji}, \citenamefont
  {Mazets}, \citenamefont {Rauer},\ and\ \citenamefont
  {Schmiedmayer}}]{cataldini2021emergent}%
  \BibitemOpen
  \bibfield  {author} {\bibinfo {author} {\bibfnamefont {F.}~\bibnamefont
  {Cataldini}}, \bibinfo {author} {\bibfnamefont {F.}~\bibnamefont
  {M{\o}ller}}, \bibinfo {author} {\bibfnamefont {M.}~\bibnamefont {Tajik}},
  \bibinfo {author} {\bibfnamefont {J.}~\bibnamefont {Sabino}}, \bibinfo
  {author} {\bibfnamefont {T.}~\bibnamefont {Schweigler}}, \bibinfo {author}
  {\bibfnamefont {S.-C.}\ \bibnamefont {Ji}}, \bibinfo {author} {\bibfnamefont
  {I.}~\bibnamefont {Mazets}}, \bibinfo {author} {\bibfnamefont
  {B.}~\bibnamefont {Rauer}},\ and\ \bibinfo {author} {\bibfnamefont
  {J.}~\bibnamefont {Schmiedmayer}},\ }\Eprint
  {https://arxiv.org/abs/arXiv:2111.13647} {arXiv:2111.13647}  (\bibinfo {year}
  {2021})\BibitemShut {NoStop}%
\bibitem [{\citenamefont {Rigol}\ and\ \citenamefont
  {Muramatsu}(2005)}]{rigol2005fermionization}%
  \BibitemOpen
  \bibfield  {author} {\bibinfo {author} {\bibfnamefont {M.}~\bibnamefont
  {Rigol}}\ and\ \bibinfo {author} {\bibfnamefont {A.}~\bibnamefont
  {Muramatsu}},\ }\href {https://doi.org/10.1103/PhysRevLett.94.240403}
  {\bibfield  {journal} {\bibinfo  {journal} {Phys. Rev. Lett.}\ }\textbf
  {\bibinfo {volume} {94}},\ \bibinfo {pages} {240403} (\bibinfo {year}
  {2005})}\BibitemShut {NoStop}%
\bibitem [{\citenamefont {Wilson}\ \emph {et~al.}(2020)\citenamefont {Wilson},
  \citenamefont {Malvania}, \citenamefont {Le}, \citenamefont {Zhang},
  \citenamefont {Rigol},\ and\ \citenamefont {Weiss}}]{wilson2020observation}%
  \BibitemOpen
  \bibfield  {author} {\bibinfo {author} {\bibfnamefont {J.~M.}\ \bibnamefont
  {Wilson}}, \bibinfo {author} {\bibfnamefont {N.}~\bibnamefont {Malvania}},
  \bibinfo {author} {\bibfnamefont {Y.}~\bibnamefont {Le}}, \bibinfo {author}
  {\bibfnamefont {Y.}~\bibnamefont {Zhang}}, \bibinfo {author} {\bibfnamefont
  {M.}~\bibnamefont {Rigol}},\ and\ \bibinfo {author} {\bibfnamefont {D.~S.}\
  \bibnamefont {Weiss}},\ }\href {https://doi.org/10.1126/science.aaz0242}
  {\bibfield  {journal} {\bibinfo  {journal} {Science}\ }\textbf {\bibinfo
  {volume} {367}},\ \bibinfo {pages} {1461} (\bibinfo {year}
  {2020})}\BibitemShut {NoStop}%
\bibitem [{\citenamefont {Bolech}\ \emph {et~al.}(2012)\citenamefont {Bolech},
  \citenamefont {Heidrich-Meisner}, \citenamefont {Langer}, \citenamefont
  {McCulloch}, \citenamefont {Orso},\ and\ \citenamefont
  {Rigol}}]{bolech2012long}%
  \BibitemOpen
  \bibfield  {author} {\bibinfo {author} {\bibfnamefont {C.~J.}\ \bibnamefont
  {Bolech}}, \bibinfo {author} {\bibfnamefont {F.}~\bibnamefont
  {Heidrich-Meisner}}, \bibinfo {author} {\bibfnamefont {S.}~\bibnamefont
  {Langer}}, \bibinfo {author} {\bibfnamefont {I.~P.}\ \bibnamefont
  {McCulloch}}, \bibinfo {author} {\bibfnamefont {G.}~\bibnamefont {Orso}},\
  and\ \bibinfo {author} {\bibfnamefont {M.}~\bibnamefont {Rigol}},\ }\href
  {https://doi.org/10.1103/PhysRevLett.109.110602} {\bibfield  {journal}
  {\bibinfo  {journal} {Phys. Rev. Lett.}\ }\textbf {\bibinfo {volume} {109}},\
  \bibinfo {pages} {110602} (\bibinfo {year} {2012})}\BibitemShut {NoStop}%
\end{thebibliography}%

\onecolumngrid
\newpage

\newcounter{equationSM}
\newcounter{figureSM}
\newcounter{tableSM}
\stepcounter{equationSM}
\setcounter{equation}{0}
\setcounter{figure}{0}
\setcounter{table}{0}
\setcounter{section}{0}
\makeatletter
\renewcommand{\theequation}{\textsc{sm}-\arabic{equation}}
\renewcommand{\thefigure}{\textsc{sm}-\arabic{figure}}
\renewcommand{\thetable}{\textsc{sm}-\arabic{table}}

\begin{center}
{\large{\bf Supplemental Material for\\
``Exact Solution of the BEC-to-BCS Quench in One Dimension"}}
\end{center}
Here we report some useful information complementing the main text. In particular
\begin{itemize}
    \item[-] In Sec.~\ref{sec:BA} we briefly review the Bethe Ansatz solution of the Gaudin-Yang model. 
    \item[-] In Sec.~\ref{sec:contOV} we explicitly compute the overlap between the Bethe states and the BEC state~(1). 
    \item[-] In Sec.~\ref{sec:QA} we present the explicit form of the quench action for quenches from the BEC state~(1).
    \item[-] In Sec.~\ref{sec:FH} we derive a formula for $g_2(\infty)$ using Feynman-Hellmann theorem. 
\end{itemize}

\section{Bethe Ansatz Treatment of the Gaudin-Yang model}
\label{sec:BA}

As shown by Gaudin and Yang~\cite{gaudin1967systeme, yang1967some}, the eigenstates of the Hamiltonian (2) can be constructed using coordinate Bethe ansatz for both repulsive, $c>0$, and attractive, $c<0$, interactions~\cite{Takahashi}. In our notations they read as 
\begin{eqnarray}
\label{eq:bethestates}
\!\!\!\!\ket{\boldsymbol k, \boldsymbol \lambda} \!=\!\!\! \sum_{\sigma_j=\pm}\int_{\mathcal D_N}\!\!\!\!\!\!\!{\rm d}{\boldsymbol x}\,\,\psi_{\boldsymbol k, \boldsymbol \lambda}(\vec{x},\boldsymbol{\sigma})\Psi^{\dag}_{\sigma_1}(x_1)\ldots\Psi^{\dag}_{\sigma_N}(x_N)\ket{0}\!.
\end{eqnarray}
Here we introduced the region 
\be
\mathcal D_N = \{ {\boldsymbol x}\in \mathbb R^N,\,\,  0\leq x_1 < \ldots < x_N\leq L\},
\ee
the vacuum state $\ket{0}$ for the fermions, and the wave function
\begin{align}
\label{eq:bethewf}
&\!\!\psi_{\boldsymbol k, \boldsymbol \lambda}(\vec{x},\boldsymbol{\sigma}) =\sum_{P\in \mathcal{S}_N}(-1)^Pe^{i\sum_j k_{P_j} x_j}\varphi_{\boldsymbol \lambda, P}(\boldsymbol{\sigma}),
\end{align}
with 
\begin{align}
&\!\!\varphi_{\boldsymbol \lambda, P}(\boldsymbol{\sigma})=\!\!\!\sum_{Q\in S_M}\prod_{Q_\alpha <Q_\beta}\!\!\!\!\frac{\lambda_{Q_\alpha}\!\!\!-\!\lambda_{Q_\beta}\!-i c}{\lambda_{Q_\alpha}\!\!\!-\!\lambda_{Q_\beta}}\prod^M_{l=1}F_P(\lambda_{Q_l},y_l),\\
&\!\! F_{P}(\lambda,y)=\frac{ic}{\lambda-k_{P_{y}}+i c/2}\prod_{j=1}^{y-1}\frac{\lambda-k_{P_{j}}-i c/2}{\lambda-k_{P_{j}}+i c/2}\,.
\end{align}
The parameters $\boldsymbol  k = \{k_i\}_{i=1}^\mathcal{N}$ and $\boldsymbol  \lambda = \{\lambda_\beta\}_{\beta=1}^M$ are known as ``rapidities" and completely specify the spectrum of $H_{\rm GY}$, as well as that of its local conservation laws. Specifically, the energy of the state $\ket{\boldsymbol k, \boldsymbol \lambda}$ is given by 
\be
E_{\boldsymbol k, \boldsymbol \lambda}=\sum_{j=1}^N k_j^2\,.
\ee 
The possible values that the rapidities can take are obtained by solving the so called Bethe equations 
\begin{align}
\label{eq:appBethe1}
\prod_{\alpha=1}^M\frac{\lambda_\alpha-{k_j}-ic/2}{\lambda_\alpha-{k_j}+ic/2}&=e^{i k_j {L} },\\
\prod_{i=1}^N\frac{\lambda_\alpha-{k_i}-ic/2}{\lambda_\alpha-{k_i}+ic/2}&=\prod_{\beta\neq\alpha}^M\frac{\lambda_\alpha-\lambda_\beta-ic}{\lambda_\alpha-\lambda_\beta+ic}.
\label{eq:appBethe2}
\end{align}

For large volumes $L$ and finite numbers of particles the solutions of the Bethe equations acquire a simple structure: each solution can be constructed combining a number of basic building blocks. These building blocks, called strings, are formed by rapidities arranged in regular patterns in the complex plane. The morphology of the strings is different in the repulsive and attractive regimes. Specifically, in the repulsive regime the $k$ rapidities are always real while, up to exponentially small corrections in $L$, $\lambda$ rapidities can form ``$\lambda$-strings" of the form   
\begin{eqnarray}
\lambda^{n,j}_\alpha=\lambda^n_\alpha+i(n+1-2j)c/2, \qquad \lambda^n_\alpha\in\mathbb R.
\end{eqnarray}
In the attractive regime, instead, there can also be complex $k$ rapidities forming ``$k-\lambda$ strings" 
\begin{eqnarray}
k^\alpha_1=\lambda^\alpha+ic/2,\qquad k_2^\alpha=\lambda^\alpha-ic/2.
\end{eqnarray}
A solution of this type describes an eigenstate containing a bound state of two particles of opposite spin.

Assuming that the description of eigenstates in terms of strings continues to hold also in the thermodynamic limit (this assumption is often referred to as ``string hypothesis"~\cite{Takahashi}) one has that states are characterised by a large number of strings and the real parts of their rapidities densely cover the real line. In this limit it is more convenient to use distributions of rapidities to specify an eigenstate. In particular, one introduces two distributions for each string type and length to describe ``occupied" and ``empty" rapidities. In our case this means that in the attractive case we have to introduce the distributions $\rho(k), \rho^h(k)$ for real rapidities and $\sigma_n(\lambda), \sigma_n^h(\lambda)$ for $\lambda$ strings, while in the attractive case we also have to add $\sigma_n(\lambda), \sigma_n^h(\lambda)$, describing the $k-\lambda$ strings. These distributions are not independent, as a consequence of the Bethe equations they are coupled together. In particular in the repulsive regime they fulfil~\cite{Takahashi}
\begin{eqnarray}
\!\!\!\!\!\!\!\!\!\rho(x)+\rho^h(x)&=&\frac{1}{2\pi}-s*\sigma^h_1(x)+R*\rho(x)\label{eq:BTrep1}\\
\!\!\!\!\!\!\!\!\!\!\sigma_n(x)+\sigma^h_n(x)&=&s*[\sigma^h_{n+1}\!+\!\sigma^h_{n-1}](x)+\delta_{n,1}s*\rho(x),\label{eq:BTrep2}
\end{eqnarray}
where
\be
f\ast g(x) =  \int {\rm d}x f(x-y)g(y)
\ee
denotes the convolution and we introduced the functions 
\be
\label{eq:sR}
s(x)=\frac{1}{2c}\sech\left(\frac{\pi x}{c}\right), \qquad R(x) = a_1 \ast s(x),
\ee
with 
\be
a_n(x)= \frac{1}{\pi}\frac{n|c|}{(nc/2)^2+x^2}. 
\ee
Instead, in the attractive regime we have 
\begin{eqnarray}
\!\!\!\!\!\!\!\!\!\!\!\!\!\rho(x)+\rho^h(x)\!&=&\!\frac{1}{2\pi}\!-\!s*\sigma^h_1(x)\!-\!R*\rho(x)-\!a_1\!*\!\tilde\sigma\!(x),\label{eq:BTatt1}\\
\!\!\!\!\!\!\!\!\!\!\!\!\!\!\sigma_n(x)\!+\!\sigma^h_n(x)\!&=&\!s*[\sigma^h_{n+1}\!+\!\sigma^h_{n-1}](x)+\delta_{n,1}s*\rho(x),\label{eq:BTatt2}\\
\!\!\!\!\!\!\!\!\!\!\!\!\!\tilde\sigma(x)\!+\!\tilde{\sigma}^{ h}(x)\!&=&\!\frac{1}{\pi}\!-\!a_2\!*\!\tilde\sigma(x)\!-\!a_1\!*\!\rho(x).\label{eq:BTatt3}
\end{eqnarray}

\section{Continuum Limit of the Overlaps}
\label{sec:contOV}
Let us consider the continuum limit of the overlaps with Bethe states. It is shown in Ref.~\cite{rylands2022integrable} that the overlaps 
\be
\frac{|{}_{\rm H}\mel*{\boldsymbol k, \boldsymbol \lambda}{(\eta^+)^{L_{\rm lat}-2\mathcal N}}{\tilde \Psi_{0,\rm lat}}|^2}{{}_{\rm H}\braket{\boldsymbol k, \boldsymbol \lambda}{\boldsymbol k, \boldsymbol \lambda}_{\rm H}},
\ee
where $\ket{\boldsymbol k, \boldsymbol \lambda}_{\rm H}$ are parity invariant Bethe states of Hubbard, read as 
\begin{align}
\label{eq:overlapH}
&\delta_{N, 2\mathcal N} \delta_{M,\mathcal N}\prod_{l=1}^{\mathcal N}\frac{\sin{k_l}^2[\cos{k_l}+U/2]^2}{\sin{k_l}^2+(U/2)^2}\prod_{\gamma=1}^{\mathcal N/2}\frac{U^4}{\lambda_\gamma^{2}{(\lambda_\gamma^2+(U/2)^2)}}\frac{\text{det}[\tilde G^+]}{\text{det}[\tilde G^-]},
\end{align}
Here $\tilde G^\pm$ denote the lattice Gaudin matrices with elements 
\begin{align} \label{UGaudinMatrixH}
&\tilde G^\pm_{i,j}=\delta_{ij}\left[\frac{L_{\rm lat}}{\cos{(k_j)}}+\sum_{\alpha=1}^{\mathcal N/2}\phi^\pm_1(\lambda_\alpha, \sin k_j)\right], \\
&\tilde G^\pm_{i,\mathcal N+\gamma}\!\!= \tilde G_{\mathcal N+\gamma,i}=-\phi^\pm_1(\lambda_\gamma-\sin k_j), \\
&\tilde G^\pm_{\mathcal N+\gamma,\mathcal N+\beta}\!\!= a^\pm_2(\lambda_\gamma,\lambda_\beta) -\delta_{\gamma \beta} \sum_{\alpha=1}^{\mathcal N/2}\phi^\pm_2(\lambda_\alpha,\lambda_\alpha) + \delta_{\gamma \beta}\sum_{j=1}^{\mathcal N}\phi^\pm_1(\lambda_\gamma,\sin k_j),
\end{align}
where $i,j\in[1,\mathcal N]$, $\gamma,\beta\in[1,\mathcal N/2]$ and finally
\be
\phi^\pm_n(\lambda,\mu)=2\pi (a_n(\lambda-\mu)\pm a_n(\lambda+\mu)).
\ee
Moreover, the rapidities are quantized according to the Bethe equations  
\begin{eqnarray}\label{eq:Bethe1H}
\prod_{\gamma}^M\frac{\lambda_\gamma-\sin{k_j}-iU/2}{\lambda_\gamma-\sin{k_j}+iU/2}&=&e^{-i k_j L_{\rm lat}},\\\label{eq:Bethe2H}
\prod_{i=1}^N\frac{\lambda_\gamma-\sin{k_i}-iU/2}{\lambda_\gamma-\sin{k_i}+iU/2}&=&\prod_{\beta\neq\gamma}^M\frac{\lambda_\gamma-\lambda_\beta-iU}{\lambda_\gamma-\lambda_\beta+iU}.
\end{eqnarray}
Recalling that the Bethe equations (5)--(6) emerge as the continuum limit of \eqref{eq:Bethe1H}--\eqref{eq:Bethe2H} upon rescaling 
\be
k\to a k, \qquad \lambda\to a \lambda
\ee
before taking the limit , we obtain the following form for the overlaps in Gaudin--Yang 
\begin{align}
\label{eq:overlapGY}
\frac{|\braket*{\boldsymbol k, \boldsymbol \lambda}{\Psi_{0}}|^2}{\braket{\boldsymbol k, \boldsymbol \lambda}{\boldsymbol k, \boldsymbol \lambda}} = &\delta_{N,2\mathcal N} \delta_{M,\mathcal N}\prod_{l=1}^{\mathcal N}\frac{{k_l}^2}{{k_l}^2+c^2/4}\prod_{\gamma=1}^{\mathcal N/2}\frac{c^4}{\lambda_\gamma^{2}{(\lambda_\gamma^2+c^2/4)}}\frac{\text{det}[G^+]}{\text{det}[G^-]},
\end{align}
and the continuum Gaudin matrices 
\begin{align} \label{UGaudinMatrix}
& G^\pm_{i,j}=\delta_{ij}\left[{L}+\sum_{\alpha=1}^{\mathcal N/2}\phi^\pm_1(\lambda_\alpha, k_j)\right], \\
& G^\pm_{i,\mathcal N+\gamma}\!\!=  G_{\mathcal N+\gamma,i}=-\phi^\pm_1(\lambda_\gamma- k_j), \\
& G^\pm_{\mathcal N+\gamma,\mathcal N+\beta}\!\!= \phi^\pm_2(\lambda_\gamma,\lambda_\beta) -\delta_{\gamma \beta} \sum_{\alpha=1}^{\mathcal N/2}\phi^\pm_2(\lambda_\alpha,\lambda_\alpha) + \delta_{\gamma \beta}\sum_{j=1}^{\mathcal N}\phi^\pm_1(\lambda_\gamma, k_j).
\end{align}

\section{The Quench Action for a quench from the BEC state}
\label{sec:QA}
The quench action is given by 
\be
\mathcal A[\rho,\sigma_n,\tilde\sigma]=- \lim_{\rm th}\frac{2}{L} \log\! \abs{\!\braket*{\Psi_0}{\boldsymbol k, \boldsymbol \mu}}-\mathcal S[\rho,\sigma_n,\tilde\sigma],
\label{eq:appQA}
\ee
where the first term comes from the extensive part of overlap between the initial state and the Bethe states and the second is the Yang-Yang entropy which counts the number of mircoscopic states corresponding to a single set of distributions.  For the attractive case the first term contains part form due real $k$,  $\lambda$  strings and $k-\lambda$ strings.  Using \eqref{eq:overlapGY} along with the fact that the ratio of determinants contributes only non extensive corrections we find 
\begin{eqnarray}\nonumber
\mathcal A[\rho,\sigma_n]&=&\int_0^\infty {\rm d}k\,\rho(k) h(k)+\int_0^\infty \!\!{\rm d}\lambda\, \tilde\sigma(\lambda)\tilde h(\lambda)+\sum_{n=1}^\infty\int_0^\infty {\rm d}\lambda \sigma_n(\lambda)h_n(\lambda)\\\nonumber
&&-\frac{1}{2}\int_{-\infty}^\infty  {\rm d}k\Big\{(\rho(k)+\rho^h(k))\log(\rho(k)+\rho^h(k))-\rho(k)\log\rho(k)-\rho^h(k)\log\rho^h(k)\Big\}\\\nonumber
&&-\frac{1}{2}\sum_n^\infty\int_{-\infty}^\infty \!\!{\rm d}\lambda \Big\{( \sigma_n(\lambda)+\sigma_n^h(\lambda))\log(\sigma_n(\lambda)+\sigma^h_n(\lambda))-\sigma_n(\lambda)\log\sigma_n(\lambda)-\sigma^h_n(\lambda)\log\sigma^h_n(\lambda)\Big\}\\\label{QuenchActionU}
&&-\frac{1}{2}\int_{-\infty}^\infty \!\!{\rm d}\lambda \Big\{(\tilde{\sigma}(\lambda)+\tilde{\sigma}^h(\lambda))\log(\tilde{\sigma}(\lambda)+\tilde{\sigma}^h(\lambda))-\tilde{\sigma}(\lambda)\log\tilde{\sigma}(\lambda)-\tilde{\sigma}^h(\lambda)\log{\tilde{\sigma}}^h(\lambda)\Big\}.
\end{eqnarray}
where in the first line the overlap for the real rapidities is given by 
\begin{align}
h(x)=f_1(x)-f_0(x),~~ f_n(x)=\log{\left[\left(x/c\right)^2+\left(n/2\right)^2\right]}
\end{align}
while those of the strings are 
\begin{eqnarray}
h_n(x)&=&\sum_{j=1}^nf_1\left(x+i(n+1-2j)c/2\right)+f_0\left(x+i(n+1-2j)c/2\right),\\~~
\tilde h(x)&=&h(x+ic/2)+h(x-ic/2)+h_1(x).
\end{eqnarray}
 The remaining lines are half the Yang-Yang entropy where the half comes from the fact that only parity invariant states have non zero overlap with the initial state.  The repulsive case can be obtained from this by dropping all the $k-\lambda$ string terms i.e. all those involving $\tilde{\sigma}$ and $\tilde{\sigma}^h$.

Varying the action with respect to the distributions leads to a set of integral equations for the ratios
\be
\zeta(x)=\frac{\rho^h(x)}{\rho(x)},\,\,\eta_n(x)=\frac{\sigma_n^h(x)}{\sigma_n(x)},\,\, \tilde\eta(x)=\frac{{\tilde \sigma}^{h}(x)}{\tilde\sigma(x)}.
\ee
Explicitly these are 
\begin{eqnarray}\label{eq:CoupledSaddle1}
\log{\zeta(k)}&=&h(k)+a_1*\log{[1+\tilde{\eta}^{-1}](k)}-\sum_{n=1}^\infty a_n*\log{[1+\eta_n^{-1}](k)}\\\label{eq:CoupledSaddle2}
\log{\tilde{\eta}}(\lambda)&=& \tilde{h}(\lambda)+a_2*\log{[1+\tilde{\eta}^{-1}](\lambda)}+a_1*\log{[1+\zeta^{-1}](\lambda)}\\\label{eq:CoupledSaddle3}
\log{\eta_n(\lambda)}&=&h_n(\lambda)+a_n*\log{[1+\zeta^{-1}](\lambda)}+\sum_{m=1}^\infty T_{mn}*\log{[1+\eta_m^{-1}](\lambda)}
\end{eqnarray}
where 
\begin{eqnarray}
T_{mn}(\lambda)=\begin{cases}a_{|n-m|}(\lambda)+2a_{|n-m|+2}(\lambda)+\dots+2a_{n+m-2}(\lambda)+a_{n+m}(\lambda)~&\text{for}~~n\neq m\\
2a_2(\lambda)+2a_4(\lambda)+\dots +2a_n(\lambda)~&\text{for}~~n= m\\
\end{cases}.
\end{eqnarray}
By using the identity
\begin{equation}
\label{eq:Tinverse}
\sum_{n}[\delta+T]_{mn}^{-1}*g_n(x) =g_m(x)-s*[g_{m+1}(x)+g_{m-1}(x)],
\end{equation} 
for some functions $g_m$, we can rearrange \eqref{eq:CoupledSaddle1}-\eqref{eq:CoupledSaddle3} to the form
\begin{align}
\log{\zeta(k)}=& \,\log{\left[\coth^2{(\pi k/2c)}\right]}\!-\!s*\log{[1+\eta_1]}(k)+s*\log{[1+\tilde\eta]}(k),\label{eq:attSadle1}\\
\log \eta_n(\lambda)=& \,\log{\left[\tanh^2{(\pi \lambda/2c)}\right]}+\delta_{n,1}s*\log{[1+\zeta^{-1}]}(\lambda)+s*\log[1+\eta_{n+1}][1+\eta_{n-1}](\lambda),\label{eq:attSadle2}\\
\log{\tilde\eta(\lambda)}=& \,\log{\frac{\lambda^4(\lambda^2+c^2)}{\mu^2(\lambda^2+c^2/4)}}+a_1*\log{[1+\zeta^{-1}]}(\lambda)+a_2*\log{[1+\tilde\eta^{-1}]}(\lambda).
\label{eq:attSadle3}
\end{align}
where $\mu$ is a Lagrange multiplier used to fix the particle density and which we determine numerically to be $\mu^2=d$.

\section{Feynman-Hellmann}
\label{sec:FH}
Noting that 
\be
\int_0^L {\rm d}x\,\Psi^\dag_+(x)\Psi_+(x)\Psi^\dag_-(x)\Psi_-(x) = \frac{1}{2}\frac{{\rm d}}{{\rm d} c} H_{\rm GY},
\ee
the expectation value of Eq. (28) can be evaluated using the Feynman-Hellmann theorem. Indeed we have 
\be
g_2(\infty) = \frac{1}{2L}\frac{{\rm d}}{{\rm d} c} {E}_{\rm sp},
\ee
where here ${E}_{\rm sp}$ is the energy of the representative state $\ket{\Phi}$. Computing the derivative and then taking the thermodynamic limit we find an expression for $g_2(\infty)$ in terms of a set of integral equations~\cite{mestyan2014short, piroli2016quantum, piroli2016multiparticle, bertini2016quantum, rylands2022integrable}. Specifically, in the repulsive case we have 
\be
g_2(\infty) =\int dk\, k\,\omega(k),
\ee
where the function $\omega(k)$ fulfils 
\begin{align}
\left[1+\zeta(k)\right]\omega(k)=& R*\omega(k)-s*\eta_1 \mu_1(k)+\mathcal{R}*\rho(k)+f*[\rho+\rho^h](k)\label{eq:FHrep1}\\
\left[1+\eta_n(\lambda)\right]\mu_n(\lambda)&= s*[\eta_{n+1}\mu_{n+1}+\eta_{n-1}\mu_{n-1}](\lambda)+f\!*\![\sigma_n+\sigma_n^h](\lambda)\!+\!\delta_{n,1}s*\omega(\lambda),\label{eq:FHrep2}
\end{align}
where $\mu_n(\lambda)$ are auxiliary functions (determined by solving the system \eqref{eq:FHrep1}--\eqref{eq:FHrep2}) and we introduced
\be
f(x)=\frac{1}{2c}\csch\left(\frac{\pi x}{c}\right), \qquad \mathcal{R}(x)=s*\mathfrak{a}_1(x), 
\ee
with 
\be
\mathfrak{a}_n(x)=\frac{x}{c} a_n(x).
\ee
Instead, in the attractive case we find
\be
g_2(\infty)= \int dk\, k\, \omega(k)+2\int d\lambda \,\lambda\,\tilde \mu(\lambda)-\frac{c}{2}\int d\lambda \tilde\sigma(\lambda),
\ee
where we have 
\begin{align}
&\left[1+\zeta(k)\right]\omega(k)= - R*\omega(k)-s*\eta_1 \mu_1(k)-\mathcal{R}*\rho(k)+f*[\rho+\rho^h](k),\label{eq:FHatt1}\\
&\left[1+\eta_n(\lambda)\right]\mu_n(\lambda)= s*[\eta_{n+1}\mu_{n+1}+\eta_{n-1}\mu_{n-1}](\lambda)+f\!*\![\sigma_n+\sigma_n^h](\lambda)\!+\!\delta_{n,1}s*\omega(\lambda),\label{eq:FHatt2}\\
&\left[1+\tilde\eta(\lambda)\right]\tilde \mu(\lambda)=-a_1*\omega(\lambda)-a_2*\tilde \mu(\lambda)-\mathfrak{a}_1*\rho(\lambda)-\mathfrak{a}_2*\tilde\sigma(\lambda)\,.\label{eq:FHatt3}
\end{align}

\end{document}